\documentclass[11pt,a4paper]{article}
\pdfoutput=1
\usepackage{jcappub}
\usepackage[english]{babel}
\usepackage{amsfonts,amsmath,amssymb,amsthm,mathtools}
\usepackage{comment,enumerate,footnote,graphicx,subfloat,relsize}
\usepackage{array,tabularx,tabu,multirow,framed,afterpage}
\usepackage[usenames,dvipsnames]{xcolor}
\numberwithin{equation}{section}
\usepackage{cleveref}
\crefformat{pluralequation}{#2\black{eqs.~(}#1\black{)}#3}
\Crefformat{pluralequation}{#2\black{Equations~(}#1\black{)}#3}
\crefformat{pluralfigure}{#2\black{figs.~}#1#3}
\Crefformat{pluralfigure}{#2\black{Figures~}#1#3}

\usepackage{tikz}
\usetikzlibrary{arrows,shapes}
\usetikzlibrary{trees,patterns}
\usetikzlibrary{matrix,arrows} 				
\usetikzlibrary{positioning}				  
\usetikzlibrary{calc,through}				  
\usetikzlibrary{decorations.pathreplacing}  
\usepackage{pgffor}							

\usetikzlibrary{decorations.pathmorphing}	
\usetikzlibrary{decorations.markings}
\tikzset{
	>=stealth', 
    vector/.style={decorate, decoration={snake}, draw},
	provector/.style={decorate, decoration={snake,amplitude=2.5pt}, draw},
	antivector/.style={decorate, decoration={snake,amplitude=-2.5pt}, draw},
	bigvector/.style={decorate, decoration={snake,amplitude=4pt}, draw},
    fermion/.style={draw=black, postaction={decorate},
        decoration={markings,mark=at position .55 with {\arrow[draw=black]{>}}}},
    fermionbar/.style={draw=black, postaction={decorate},
        decoration={markings,mark=at position .55 with {\arrow[draw=black]{<}}}},
    fermionnoarrow/.style={draw=black},
    gluon/.style={decorate, draw=black,
        decoration={coil,amplitude=4pt, segment length=5pt}},
    scalar/.style={dashed,draw=black, postaction={decorate},
        decoration={markings,mark=at position .55 with {\arrow[draw=black]{>}}}},
    scalarbar/.style={dashed,draw=black, postaction={decorate},
        decoration={markings,mark=at position .55 with {\arrow[draw=black]{<}}}},
    scalarnoarrow/.style={dashed,draw=black},
    momentum/.style={draw=black, postaction={decorate},
        decoration={markings,mark=at position 1 with {\arrow[draw=black]{>}}}},
    antimomentum/.style={draw=black, postaction={decorate},
        decoration={markings,mark=at position 0.1 with {\arrow[draw=black]{<}}}}
}

\tikzstyle{block} = [draw, rectangle, 
    minimum height=3em, minimum width=6em]

\newcommand{\nc}{\newcommand}
\nc{\pd}{\partial}
\nc{\bea}{\begin{eqnarray}}
\nc{\eea}{\end{eqnarray}}
\nc{\bal}{\begin{alignedat}}
\nc{\eal}{\end{alignedat}}
\nc{\beq}{\begin{equation}}
\nc{\eeq}{\end{equation}}
\nc{\bit}{\begin{itemize}}
\nc{\eit}{\end{itemize}}
\nc{\benu}{\begin{enumerate}}
\nc{\eenu}{\end{enumerate}}
\nc{\bdes}{\begin{description}}
\nc{\edes}{\end{description}}
\nc{\bma}{\begin{pmatrix}}
\nc{\ema}{\end{pmatrix}}

\newcommand{\black}[1]	{{\color{black} 	#1}}

\nc{\nn}{\nonumber}
\nc{\hc}{\text{h.c.}}
\nc{\cc}{\text{c.c.}}

\nc{\slashed}[1]{{#1}\hspace{-2mm}/}

\nc{\abs}[1]{\left| #1 \right|}
\def\[{\left[}
\def\]{\right]}
\def\({\left(}
\def\){\right)}
\def\<{\langle}
\def\>{\rangle}

\def\g5{\gamma_{5}}

\def \eV{{\rm eV}}
\def\keV{{\rm keV}}
\def\MeV{{\rm MeV}}
\def\GeV{{\rm GeV}}
\def\TeV{{\rm TeV}}

\def \pc{{\rm pc}}

\def\a{\alpha}
\def\b{\beta}
\def\g{\gamma}
\def\d{\delta}

\def\z{\zeta}

\def\l{\lambda}
\def\m{\mu}

\def\p{\pi}

\def\s{\sigma}

\def\vf{\varphi}

\def\G{\Gamma}

\def\aD			{\alpha_{\mathsmaller{D}}}
\def\aEM		{\alpha_{\mathsmaller{\rm EM}}}
\def\ann		{{\rm ann}}
\def\BSF		{\mathsmaller{\rm BSF}}
\def\etaB		{\eta_{\mathsmaller{B}}}
\def\etaD		{\eta_{\mathsmaller{D}}}
\def\FO			{{\mathsmaller{\rm FO}}}
\def\FD			{F_{\mathsmaller{D}}}
\def\FY			{F_{\mathsmaller{Y}}}
\def\inel		{{\rm inel}}
\def\mVD		{m_{\mathsmaller{V_D}}}
\def\MDM		{M_{\mathsmaller{\rm DM}}}
\def\MPl		{M_{\mathsmaller{\rm Pl}}}
\def\gX			{g_{\mathsmaller{X}}}
\def\gD			{g_{\mathsmaller{D}}}
\def\gSM		{g_{\mathsmaller{\rm SM}}}
\def\gtildeD	{\tilde{g}_{\mathsmaller{D}}}
\def\gtildeSM	{\tilde{g}_{\mathsmaller{\rm SM}}}
\def\geff		{g_{\rm eff}}
\def\heff		{h_{\rm eff}}
\def\OmegaB		{\Omega_{\mathsmaller{\rm B}}}
\def\OmegaDM	{\Omega_{\mathsmaller{\rm DM}}}
\def\rinf		{r_\infty}

\def\sD			{s_{\mathsmaller{D}}}
\def\sSM		{s_{\mathsmaller{\rm SM}}}
\def\sym		{{\rm sym}}
\def\TD			{T_{\mathsmaller{D}}}
\def\TSM		{T_{\mathsmaller{\rm SM}}}
\def\uni		{{\rm uni}}
\def\VD			{V_{\mathsmaller{D}}}
\def\vEW		{v_{\mathsmaller{\rm EW}}}
\def\vrel		{v_{\rm rel}}
\def\xiD		{\xi_{\mathsmaller{D}}}
\def\xD			{x_{\mathsmaller{D}}}
\def\zD			{z_{\mathsmaller{D}}}

\preprint{DESY 17-034, \\ \phantom{} \hfill Nikhef 2017-009}

\title{\LARGE
Asymmetric thermal-relic dark matter: \\
Sommerfeld-enhanced freeze-out, annihilation signals and unitarity bounds}

\author[a]{\large Iason Baldes}
\author[b,c]{\large and Kalliopi Petraki}

\affiliation[a]{DESY, Notkestra{\ss}e 85, D-22607 Hamburg, Germany}
\affiliation[b]{Laboratoire de Physique Th\'eorique et Hautes Energies (LPTHE), UMR 7589 CNRS \& UPMC, 4 Place Jussieu, F-75252, Paris, France}
\affiliation[c]{Nikhef, Science Park 105, 1098 XG Amsterdam, The Netherlands}

\emailAdd{iason.baldes@desy.de}
\emailAdd{kpetraki@lpthe.jussieu.fr}

\date{\today}

\abstract{ 
Dark matter that possesses a particle-antiparticle asymmetry and has thermalised in the early universe, requires a larger annihilation cross-section compared to symmetric dark matter, in order to deplete the dark antiparticles and account for the observed dark matter density. The  annihilation cross-section determines the residual symmetric component of dark matter, which may give rise to annihilation signals during CMB and inside haloes today. 
We consider dark matter with long-range interactions, in particular dark matter coupled to a light vector or scalar force mediator. We compute the couplings required to attain a final antiparticle-to-particle ratio after the thermal freeze-out of the annihilation processes in the early universe, and then estimate the late-time annihilation signals. We show that, due to the Sommerfeld enhancement, highly asymmetric dark matter with long-range interactions can have a significant annihilation rate, potentially larger than symmetric dark matter of the same mass with contact interactions. We discuss caveats in this estimation, relating to the formation of stable bound states. 
Finally, we consider the non-relativistic partial-wave unitarity bound on the inelastic cross-section, we discuss why it can be realised only by long-range interactions, and showcase the importance of higher partial waves in this regime of large inelasticity. We derive upper bounds on the mass of symmetric and asymmetric thermal-relic dark matter for $s$-wave and $p$-wave annihilation, and exhibit how these bounds strengthen as the dark asymmetry increases.
}

\arxivnumber{1703.00478}

\begin{document}
\maketitle

\section{Introduction \label{Sec:Intro}}

If dark matter (DM) transforms under a global U(1) symmetry that governs its low-energy interactions, it is possible that today there are unequal densities of  dark particles and dark antiparticles.  The dark particle-antiparticle asymmetry may have been related to the baryonic asymmetry of ordinary matter, via high-energy processes that occurred in the early universe, thereby providing a dynamical explanation for the similarity between the dark and the ordinary matter densities. Independently of such a connection, asymmetric DM can be a thermal relic of the primordial plasma while still having large couplings to lighter species, since its abundance cannot be depleted below the conserved excess of dark particles over antiparticles, via annihilations to these light species. Asymmetric DM thus provides a compelling cosmological scenario for large portions of the low-energy parameter space in a variety of beyond-the-Standard-Model theories, including models with new stable particles coupled to the Weak interactions of the Standard Model (WIMPs), as well as hidden-sector models~\cite {Petraki:2013wwa}.

In the asymmetric DM scenario, the efficiency of the annihilation processes in the early universe determines the relative abundance of dark particles and antiparticles today, i.e.~the residual symmetric DM component. This, in turn, determines the DM annihilation signals at late times, that could be looked for by the ongoing indirect DM searches. The DM freeze-out in the presence of a particle-antiparticle asymmetry was first considered in~\cite{Scherrer:1985zt, Griest:1986yu}, and more recently computed in greater detail and generality in~\cite{Graesser:2011wi, Iminniyaz:2011yp}. Reference~\cite{Graesser:2011wi} showed that the residual dark antiparticle-to-particle ratio decreases exponentially with the DM annihilation cross-section. It then appears reasonable that sizeable annihilation signals may be expected only for annihilation cross-sections close to that for symmetric thermal-relic DM. This is indeed valid if DM annihilates via contact-type interactions~\cite{Graesser:2011wi, Bell:2014xta, Murase:2016nwx}, i.e.~interactions that can be neglected at large distances and can be treated perturbatively.  
This type of interactions were the focus of previous investigations~\cite{Scherrer:1985zt, Griest:1986yu, Graesser:2011wi, Iminniyaz:2011yp, Bell:2014xta, Murase:2016nwx}.

In this paper, we consider asymmetric DM coupled to light force mediators. If a mediator is sufficiently light, then the interaction between DM particles manifests as long-range. More specifically, for an interaction that is described in the non-relativistic regime by a Yukawa potential, $V = - \aD \, e^{-m_{\rm med} \, r}/r$, long-range effects arise if the mediator mass is smaller than the Bohr momentum, 
\beq m_{\rm med} \lesssim \aD \MDM/2\,, \label{eq:long-range} \eeq 
where $\MDM$ is the DM mass. The long-range interaction distorts the wavefunction of the dark particle-antiparticle pairs, giving rise to the well-known Sakharov-Sommerfeld effect~\cite{SakharovEffect,Sommerfeld:1931} (in the following referred to only by ``Sommerfeld", for brevity), which enhances the DM annihilation rate at low velocities~\cite{Hisano:2002fk, Hisano:2003ec}. In addition, long-range interactions imply the existence of bound states~\cite{Pospelov:2008jd, MarchRussell:2008tu, Shepherd:2009sa, Feng:2009mn, vonHarling:2014kha, Petraki:2015hla, Liew:2016hqo, Petraki:2016cnz}, whose formation is also a Sommerfeld-enhanced process~\cite{Petraki:2015hla}. The formation of unstable particle-antiparticle bound states, and their subsequent decay contributes to the overall DM annihilation rate. These non-perturbative phenomena, the Sommerfeld effect and the formation of bound states, reduce the couplings required to attain the observed DM density via thermal freeze-out in the early universe~\cite{Hisano:2003ec, vonHarling:2014kha}; on the other hand, for a specified set of couplings, they enhance the late-time DM annihilation signals~\cite{Hisano:2004ds, Pospelov:2008jd, MarchRussell:2008tu, ArkaniHamed:2008qn, Belotsky:2014doa, An:2016gad, An:2016kie, Petraki:2016cnz, Bringmann:2016din, Cirelli:2016rnw}. The present work investigates the interplay between these two effects, in the context of asymmetric DM. Since the Sommerfeld enhancement depends on the coupling of DM to the light force mediator, and asymmetric DM requires stronger couplings than symmetric DM of the same mass, the implications for the phenomenology of asymmetric DM may be rather significant.\footnote{
A related computation of asymmetric freeze-out with Sommerfeld-enhanced cross-sections appeared recently in Ref.~\cite{Agrawal:2017rvu}, which focused on the energy dissipation that would take place inside halos, in an atomic DM scenario. The scope and the extent of the two studies are very different.}

We shall consider two minimal scenarios, in which DM is coupled to a massless or light vector boson, a dark photon, or to a light scalar mediator. We compute the couplings required to establish the observed DM abundance as a function of the dark asymmetry, and demonstrate the impact of the Sommerfeld effect. Using these computations, we estimate the strength of the radiative signals expected from the annihilation of the residual symmetric DM component inside haloes today. We find that \emph{highly} asymmetric DM with long-range interactions can give rise to annihilation signals that are stronger than those of symmetric DM with contact interactions, up to several orders of magnitude. This is in sharp contrast with the common expectation that asymmetric DM with antiparticle-to-particle ratio much lower than 1 yields negligible annihilation signals.\footnote{%
Note that we do not assume the dark antiparticles were re-populated after an initial phase of dark asymmetry, as in scenarios that feature dark particle-antiparticle oscillations~\cite{Cirelli:2011ac} or decays of heavier species~\cite{Hardy:2014dea}.} 
We discuss caveats to this estimate, related to the possible formation of stable bound states by asymmetric DM in the early universe.

Partial-wave unitarity sets an upper limit on inelastic cross-sections. This has been invoked to deduce the maximum mass for which thermalised DM can annihilate sufficiently in the early universe, to attain the observed density~\cite{Griest:1989wd}. Asymmetric thermal-relic DM requires more efficient annihilation than symmetric DM; the upper mass bound implied by unitarity must, thus, tighten for larger values of the DM asymmetry.\footnote{%
Unitarity of the S-matrix also plays a crucial role in constraining the generation of an asymmetry~\cite{Weinberg:1979bt, Kolb:1979qa, Dolgov:1979mz, Baldes:2014gca, Baldes:2015lka}. The focus here, however, is the subsequent annihilation of the symmetric DM component. Asymmetric DM may also remain non-thermal throughout the cosmological history, in which case the computations of the present work, including the unitarity bounds, do not apply. The possibility of non-thermal asymmetric DM is encountered, for example, in the scenario of stable $Q$-balls produced in the fragmentation of an Affleck-Dine condensate~\cite{Kusenko:1997ad,Kusenko:1997zq,Kusenko:1997si}.} 
It has been pointed out that in the non-relativistic regime, the unitarity limit on the inelastic cross-section can be realised only via long-range interactions~\cite{vonHarling:2014kha}. We expound on the pertaining arguments, and further assert that in the regime where the unitarity limit may be realised, partial waves beyond the lowest one need to be considered. 
We then employ the freeze-out calculations with Sommerfeld-enhanced cross-sections carried out in this work, to compute the unitarity bounds on the mass and the asymmetry of thermal-relic DM, for the dominant partial waves that appear in known inelastic processes.

\medskip

The rest of the paper is organised as follows. 
In \cref{Sec:AsymFO}, we review the computation of the DM relic density in the presence of a conserved particle-antiparticle asymmetry. We follow closely the analysis of Ref.~\cite{Graesser:2011wi}, and generalise it whenever necessary for our purposes. 
In \cref{Sec:SommerfeldAsymFO}, we consider asymmetric DM coupled to 
light vector and scalar bosons, and compute the couplings required to establish the observed DM density, as a function of the DM asymmetry. 
In \cref{Sec:AnnihilationSignals}, we estimate the expected indirect detection signals from the late-time annihilation of the residual symmetric DM component, and contemplate possible complications. 
In \cref{Sec:Unitarity}, we discuss and compute the bounds implied by unitarity on symmetric and asymmetric thermal-relic DM. We conclude in \cref{Sec:Concl}.

\section{Thermal freeze-out in the presence of an asymmetry \label{Sec:AsymFO}}

\subsection{The dark-sector temperature \label{Sec:DStemperature}}

The dark plasma --- the bath of dark-sector relativistic particles into which DM annihilates --- may be in general at a different temperature than photons. We will assume that at early times, the dark sector was in thermal equilibrium with the Standard Model (SM) plasma due to some unspecified high-energy interactions that decoupled at a high temperature $\tilde{T}$. Beyond this point, the SM and dark-sector temperatures, $\TSM$ and $\TD$, evolve differently. The SM-sector, dark-sector and total entropy densities are 
$\sSM = (2\pi^2/45) \, \gSM \, \TSM^3$, $\sD = (2\pi^2/45) \, \gD \, \TD^3$ and $s = \sSM + \sD$ respectively, 
where $\gSM$ and $\gD$ are the SM and dark-sector relativistic degrees of freedom, which depend on the temperatures.   Assuming conservation of co-moving entropy in each sector separately below the common temperature $\tilde{T}$, the dark-to-ordinary temperature ratio is
\beq
\tau \equiv \frac{\TD}{\TSM} 
= \(\frac{\gSM}{\gD}\)^{1/3} \(\frac{\gtildeD}{\gtildeSM}\)^{1/3} \,,
\label{eq:TD/TSM}
\eeq
where $\gtildeSM$ and $\gtildeD$ refer to the temperature $\tilde{T}$. For our purposes, it will be convenient to express the total entropy and energy densities in terms of $\TD$,
\begin{subequations}
\label{eq:EntropyEnergyDensities}
\label[pluralequation]{eqs:EntropyEnergyDensities}
\beq
s = (2\pi^2 / 45) \, \heff(\TD) \, \TD^3 \,,  \qquad
\rho = (\pi^2 / 30) \, \geff(\TD) \, \TD^4 \,,
\tag{\ref{eq:EntropyEnergyDensities}}
\label{eq:rho_s}
\eeq
where
\begin{align}
\heff &\equiv \gSM/\tau^3 + \gD  \,=\, \gD \, (1+\gtildeSM/\gtildeD) \,,
\\
\geff &\equiv \gSM/\tau^4 + \gD  \,=\, \gD \, [1+ (\gD/\gSM)^{1/3} \, (\gtildeSM/\gtildeD)^{4/3} ] \,.
\label{eq:geff_heff}
\end{align}
\end{subequations}
The Hubble parameter in the radiation dominated epoch is $H = \sqrt{4\pi^3 \geff/45} \: \TD^2/\MPl$.
Note that we do not distinguish between the entropy and the energy degrees of freedom for the SM and the dark sector, since there is no difference within the temperature range of interest.

We use the values of $\gSM$ available with the MicrOMEGAs package~\cite{Belanger:2001fz}, and assume that $\gtildeSM$ includes all the SM degrees of freedom, i.e.~that $\tilde{T}$ is larger than the temperature of the electroweak phase transition. 
In addition, we shall assume that $\tilde{T} > \MDM/3$, and take $\gtildeD = (7/8)\times 4 + 1= 4.5$  or $\gtildeD = (7/8)\times 4 +2 = 5.5$, to account for the four degrees of freedom of DM consisting of Dirac Fermions, plus a real scalar or vector force mediator respectively. We take $\gD = \gtildeD$ for $\TD \gtrsim \MDM/3$ and $\gD = \gtildeD-(7/8) \times 4$ for $\TD < \MDM/3$.

\medskip
We note that under these assumptions, a massless or very light mediator ($m_{\rm med} \lesssim \eV$) would contribute to the relativistic energy density during CMB by $\delta N_{\rm eff} \approx 0.2$, which is well within the 1$\sigma$ range of the \emph{Planck} measurement 
$N_{\rm eff} = 3.15 \pm 0.23$~\cite[\emph{Planck} TT+lowP+BAO]{Ade:2015xua}. (Accounting for massive neutrinos further relaxes the constraint on $N_{\rm eff}$~\cite[eq.~(67)]{Ade:2015xua}; see also Ref.~\cite{Vagnozzi:2017ovm} for related discussion.) The contribution to $N_{\rm eff}$ is eliminated if the mediator is somewhat massive. However, the cosmological density of a light mediator with non-zero mass could dominate the universe after the mediator became non-relativistic, thus altering the time of matter-radiation equality. Therefore light mediators must have decayed sufficiently early, either into SM particles via a portal operator or into other lighter dark-sector species, or they must have become non-relativistic after their density redshifted to a negligible amount (see e.g.~Refs.~\cite{Cirelli:2016rnw,Bringmann:2016din} for relevant considerations on a particular model).

The above cosmological considerations become less constraining in the context of a particular model if there were additional degrees of freedom coupled to the SM sector at $\tilde{T}$. The subsequent decoupling of the latter from the SM thermal bath would suppress the dark-to-ordinary temperature ratio $\tau$ and lower $\delta N_{\rm eff}$. Moreover, it is possible that the dark sector was at a lower temperature in early times due to initial conditions set by inflation (see e.g.~\cite{Adshead:2016xxj}). Of course, a lower dark-sector temperature would also affect the DM freeze-out and decrease the estimated couplings that can produce the observed DM density; the estimated annihilation cross-section at the time of freeze-out would have to be lower by approximately the same factor as the dark-sector temperature.

\subsection{Boltzmann equations}

We shall use $\xD \equiv \MDM/\TD$ as the time variable, and parametrise the thermally averaged annihilation cross-section times relative velocity as
\beq
\<\s_\ann\vrel\> \equiv \sigma_* \times F(\xD) \,.
\label{eq:sigma vrel}
\eeq
For fully perturbative $s$- and $p$-wave annihilation $F(\xD) = 1$ and $F(\xD) = \<\vrel^2\> = 6/\xD$ respectively; however, for an inelastic process well within the Sommerfeld-enhanced regime and the Coulomb approximation, $F(\xD) \propto \<1/\vrel\> \propto \xD^{1/2}$, independently of the partial wave it may be dominated by~\cite{Cassel:2009wt, vonHarling:2014kha, Petraki:2015hla}. More details on the annihilation cross-sections will be specified in \cref{Sec:SommerfeldAsymFO}. 

The DM particle and antiparticle number-to-entropy ratios, $Y^{\pm} \equiv n^{\pm}/s$, evolve according to
\begin{subequations}
\label{eq:Boltzmann_Y}
\label[pluralequation]{eqs:Boltzmann_Y}
\beq
\frac{dY^{\pm}}{d\xD} = -\frac{g_*^{1/2} \, \lambda\, F(\xD)}{\xD^2} 
\[Y^+(\xD) \, Y^-(\xD) - Y_{\rm eq}^\sym(\xD)^2\] \,,
\tag{\ref{eq:Boltzmann_Y}}
\label{eq:dY/dz}
\eeq
where 
\begin{align}
g_*^{1/2} &\equiv \frac{\heff}{\geff^{1/2}} 
\(1+ \frac{\TD}{3\heff}\frac{d\heff}{d\TD}\) \,,
\label{eq:gstar}
\\
\lambda &\equiv \sqrt{\frac{\p}{45}} \: \sigma_* \, \MDM \, \MPl \,,
\label{eq:lambda}
\\
Y_{\rm eq}^\sym (\xD) &\simeq 
\frac{90}{(2\pi)^{7/2}} \, \frac{\gX }{\heff} \, \xD^{3/2} \, e^{-\xD} \,.
\label{eq:Yeq_sym}
\end{align}
In the above, $Y_{\rm eq}^\sym$ is the equilibrium value of $Y^{\pm}$ in the absence of an asymmetry, with $\gX$ being the DM degrees of freedom. In the presence of an asymmetry, the equilibrium values of $Y^\pm$ are 
\beq
Y_{\rm eq}^{\pm} (\xD) = Y_{\rm eq}^\sym (\xD) \, e^{\pm \xiD} \,,
\label{eq:Yeq_+-}
\eeq 
\end{subequations}
where $\xiD \equiv \mu/\TD$, with $\mu$ being the equilibrim chemical potential, which evolves with $\TD$ in order to account for the conserved dark particle-antiparticle asymmetry, as we shall now see [cf.~\cref{eq:xiD}]. 

We define two asymmetry parameters, the fractional asymmetry $r$ and the dark particle-minus-antiparticle-number-to-entropy ratio $\etaD$,
\begin{align}
r(\xD) &\equiv Y^-(\xD)/Y^+(\xD) \,, \label{eq:r_def} \\
\etaD &\equiv Y^+ - Y^- \,. \label{eq:eta}
\end{align}
\Cref{eq:r_def,eq:eta} can be inverted to give $Y^+ = \etaD/(1-r)$ and $Y^- = \etaD\, r/(1-r)$. 
In an isentropically expanding universe, $\etaD$ is conserved. As long as DM is in chemical equilibrium with dark radiation, the fractional asymmetry is 
$r_{\rm eq} \equiv Y_{\rm eq}^-/Y_{\rm eq}^+ = \exp(-2 \xiD)$, 
where the equilibrium chemical potential over temperature is determined from 
\cref{eq:eta,eq:Yeq_+-} to be
\beq
\xiD (\xD) = \ln \left[ \sqrt{1+ \(\dfrac{\etaD}{2Y_{\rm eq}^{\sym \vphantom{S}} (\xD)}\)^2}
+ \dfrac{\etaD }{ 2Y_{\rm eq}^{\sym \vphantom{S}} (\xD)} \right] \,.
\label{eq:xiD}
\eeq
Ultimately, we are interested in computing the final fractional asymmetry, 
\beq \rinf \equiv \lim_{\xD \to \infty} r(\xD) \,,\eeq 
which determines the DM annihilation signals today, and the predicted DM mass.

\subsection{Dark matter mass and its maximum value}

The ratio of DM to ordinary matter relic energy densities is 
$\OmegaDM/\OmegaB = (Y_\infty^+ + Y_\infty^-) \MDM / (\etaB \, m_p)$, 
where $\etaB$ is the baryon-number-to-entropy ratio of the universe, and $m_p$ is the proton mass.  
Using \cref{eq:r_def,eq:eta}, and setting $\epsilon \equiv \etaD/\etaB$, we obtain
\beq
\MDM = \frac{m_p}{\epsilon} \frac{\OmegaDM}{\OmegaB} 
\(\frac{1-\rinf}{1+\rinf}\) \,.
\label{eq:MDM}
\eeq
For non-zero $\epsilon$, \cref{eq:MDM} implies a maximum DM mass
\beq
\MDM < M_{\rm max}(\epsilon) 
\equiv \frac{m_p}{\epsilon} \frac{\OmegaDM}{\OmegaB} 
\simeq 5~\GeV / \epsilon
\,,
\label{eq:MDM_max}
\eeq
attained in the limit $\rinf \to 0$. This would, however, require an infinitely large cross-section. Partial-wave unitarity sets an upper limit on the inelastic cross-section, and thus implies $\rinf >0$, which in turn strengthens the upper bound on the mass of (asymmetric) thermal-relic DM to $\MDM < M_{\rm uni} < M_{\rm max}$. In \cref{Sec:Unitarity}, we show how $M_{\rm uni}$ varies with the DM asymmetry.

\subsection[Final fractional asymmetry $\rinf$]{Final fractional asymmetry $\boldsymbol{\rinf}$}

From \cref{eq:dY/dz}, we find that $r$ is governed by the equation~\cite{Graesser:2011wi}
\beq
\frac{dr}{d\xD} = -\frac{\etaD \lambda g_*^{1/2} \, F(\xD)}{\xD^2} 
\[r - r_{\rm eq} \(\frac{1-r}{1-r_{\rm eq}}\)^2\] \,.
\label{eq:dr/dz}
\eeq
Soon after freeze-out, the second term in \cref{eq:dr/dz} becomes unimportant, and the evolution of $r$ is determined by the first term. The final fractional asymmetry can thus be approximated by
\beq
\rinf \simeq r_{\rm eq}^\FO  \exp[-\etaD \lambda \Phi (\aD)] \,,
\label{eq:rinf_approx}
\eeq
where
\begin{subequations}
\label{eq:Phi}
\label[pluralequation]{eqs:Phi}
\beq
\Phi (\aD) \equiv \int_{\xD^\FO}^\infty d\xD \: g_*^{1/2} \: F(\xD)/\xD^2 \,.
\tag{\ref{eq:Phi}}
\label{eq:Phi_def}
\eeq
Here, we have chosen to emphasise the possible dependence of $\Phi$ on the couplings of the theory (denoted by $\aD$) that arises in Sommerfeld-enhanced cross-sections (to be specified in \cref{Sec:SommerfeldAsymFO}). This is important for the determination of $\rinf$, as discussed below. 
In the cases of interest, the function $F(\xD)$ either decreases with $\xD$ or grows at most as $\xD^{1/2}$, therefore the integral \eqref{eq:Phi_def} is dominated by the contribution at $\xD \approx \xD^\FO$, and can be approximated as
\beq
\Phi (\aD) \simeq
\dfrac{ \sqrt{g_*^\FO} \ F(\xD^\FO)}{c_{\mathsmaller{\Phi}} \, \xD^\FO} \,,
\label{eq:Phi_approx}
\eeq
\end{subequations}
where $c_{\mathsmaller{\Phi}} \sim {\cal O}(1)$ is a numerical factor that will not appear in our final result below. Freeze-out --- the time when the densities of the DM particles and antiparticles depart from their equilibrium values --- occurs when the terms contributing to the logarithmic derivatives, $|d \ln Y^\pm / d\xD|$, of the dark particle and antiparticle densities become small, i.e.~at 
\beq 
\lambda \, \sqrt{g_*^\FO}  \, F(\xD^\FO) \, Y_{\rm eq}^\sym(\xD^\FO) \,/\, (\xD^\FO)^2  
\ \sim \ 1/c_x  \nn
\eeq 
[cf.~\cref{eq:dY/dz}], where  $c_x \sim {\cal O}(1)$. This yields the standard algebraic equation for $\xD^\FO$, which in our formalism reads
\beq
\xD^\FO + (1/2) \ln \xD^\FO -\ln F(\xD^\FO) \simeq 
\ln (c_x \, 0.15 \, \lambda \, \gX/\sqrt{g_*^\FO} )  \,.
\eeq 
Note that $\xD^\FO \approx 20-30$ is insensitive to the presence of an asymmetry~\cite{Graesser:2011wi}. Moreover, using \cref{eq:Phi_approx}, $Y_{\rm eq}^\sym$ can be re-expressed as $Y_{\rm eq}^\sym(\xD^\FO) \approx \xD^\FO / (c \lambda \Phi)$, where $c = c_{\mathsmaller{\Phi}} c_x$. From \cref{eq:xiD}, we may now estimate the fractional asymmetry at freeze-out, 
\beq 
r_{\rm eq}^\FO  \, \simeq \, e^{-2\xiD^\FO} \, \approx \,
\[
\sqrt{1 + (c\,\etaD\lambda\Phi)^2/(2\xD^\FO)^2} + 
(c\,\etaD\lambda\Phi) / (2\xD^\FO) 
\]^{-2}\,.
\label{eq:r_eq^FO}
\eeq 
For an interaction that scales as $F(\xD) \propto \xD^{-n}$, $c_x \approx c_{\mathsmaller{\Phi}} = n+1$ (see e.g.~Ref.~\cite{Kolb:1990vq});  consequently $c \approx (n+1)^2$.

Collecting \cref{eq:rinf_approx,eq:r_eq^FO}, we obtain
\beq
\rinf \simeq 
\frac{\exp(- \etaD \lambda \Phi)}
{\[\sqrt{1 + \(\dfrac{c\,\etaD\lambda\Phi}{ 2\xD^\FO}\)^2} + 
\(\dfrac{c\,\etaD\lambda\Phi}{ 2\xD^\FO}\)  \]^{2}} \,.
\label{eq:rinf_final_full}
\eeq
(This expression is more general than the expressions provided in Ref.~\cite{Graesser:2011wi}.) 
Evidently, $\rinf$ depends on the combination of parameters $\etaD \lambda \Phi$. A direct comparison with symmetric DM can be established by recasting $\etaD$ in terms of $\rinf$ using \cref{eq:MDM}, and recalling that in the symmetric DM limit, 
$\OmegaDM / \OmegaB = (2 Y_\infty^\sym \MDM) / (\etaB m_p )$
where the relic number-to-entropy ratio is 
\beq 
Y_\infty^\sym \simeq [(1+c/\xD^\FO) \, \lambda_\sym \Phi_\sym]^{-1}, 
\label{eq:Yinf_sym}
\eeq
as can be deduced from \cref{eq:dY/dz}. Then, we find
\beq
\etaD \lambda \Phi = 
\frac{2}{1+c/\xD^\FO}
\(\frac{1-\rinf}{1+\rinf}\)
\(\frac{\sigma_*}{\sigma_{*,\rm sym}}\)
\(\frac{\Phi}{\Phi_\sym}\) \,,
\label{eq:lambda_eta_Phi}
\eeq
where the subscript ``sym" refers to symmetric DM of the same mass. 
For a small asymmetry $\etaD$, we expand \cref{eq:rinf_final_full} as
\beq 
\ln (\rinf) \approx -(1+c/\xD^\FO) \etaD \lambda \Phi 
+ {\cal O}[(c/\xD^\FO)^3\etaD^3\lambda^3\Phi^3]. 
\label{eq:ln(rinf) expansion}
\eeq
Keeping only the lowest order term is a good approximation for $(c/\xD^\FO) \etaD \lambda \Phi \lesssim 1$; 
because $c/\xD^\FO \ll 1$, this range extends to very small $\rinf$. 
Then, using \cref{eq:lambda_eta_Phi}, we arrive at the result of Ref.~\cite{Graesser:2011wi},
\beq
\rinf \simeq
\exp\[-2
\(\frac{1-\rinf}{1+\rinf}\)
\(\frac{\sigma_*}{\sigma_{*,\rm sym}}\)
\(\frac{\Phi}{\Phi_\sym}\) 
\] 
\,.
\label{eq:rinf_final_approx}
\eeq
Note that using the approximation \eqref{eq:Phi_approx}, 
$(\sigma_* \, \Phi) / (\sigma_{*, \rm sym} \, \Phi_\sym) 
\simeq \sigma_\ann^\FO / \sigma_{\rm ann,\, sym}^\FO$, provided that $F(\xD)$ scales with $\xD$ in the same way, around the time of freeze-out, for the couplings corresponding to the symmetric and asymmetric cases.

\medskip

As is evident from \cref{eq:rinf_final_approx}, $\rinf$ depends exponentially on the annihilation cross-section at freeze-out. Therefore, a cross-section only somewhat larger than that required for symmetric thermal-relic DM, suffices to diminish the antiparticle density considerably~\cite{Graesser:2011wi}. 
For annihilation via fully perturbative processes, the dependence of the cross-section on the couplings of the theory and on the DM velocity (or temperature) can be factorised inside $\sigma_*$ and $F(\xD)$ respectively;  $\Phi$ depends only on $\xD^\FO$, which is insensitive to $\sigma_*$ (i.e.~the couplings of the theory), thus $\Phi \simeq \Phi_\sym$. This case was investigated in detail in Ref.~\cite{Graesser:2011wi}. 
However, for Sommerfeld-enhanced cross-sections, this factorisation is not in general possible; $\Phi$ depends on the couplings of the theory explicitly (rather than via $\xD^\FO$ only), and can differ significantly from $\Phi_\sym$. This enhances the sensitivity of $\rinf$ to the strength of the interactions, and implies that a small $\rinf$ can be attained for more modest couplings.  For an annihilation cross-section that scales around the time of freeze-out as $\sigma_\ann^\FO \propto \aD^p$, we find from \cref{eq:rinf_final_approx},
\beq 
\aD / \aD^\sym \simeq 
\[\(\frac{1+\rinf}{1-\rinf}\) \frac{\ln (1/\rinf)}{2} \]^{1/p} \,.
\label{eq:AlphaRatio}
\eeq
We investigate the effect of the Sommerfeld enhancement for specific interactions in the next section.

\section{Asymmetric freeze-out with Sommerfeld-enhanced cross-sections \label{Sec:SommerfeldAsymFO}}

We will consider two minimal cases, in which DM consists of Dirac Fermions and couples either to a light vector or scalar boson. In both cases, the interaction between dark particles and antiparticles is described in the non-relativistic regime by a static Yukawa potential, $V_Y({\bf r}) = - \aD \, e^{-m_{\rm med} \, r} / r$. 
We will perform all computations in the Coulomb limit, $m_{\rm med} \to 0$, which is a satisfactory approximation if the average momentum transfer between the interacting particles is larger than the mediator mass~(see e.g.~\cite{Petraki:2016cnz}), 
\beq m_{\rm med} \lesssim \vrel (\MDM/2) \,. \label{eq:CoulombRegime} \eeq 
The Coulomb approximation is suitable during the DM chemical decoupling in the early universe essentially in the entire range where non-perturbative effects arise, i.e.~for $m_{\rm med} \lesssim \aD (\MDM/2)$~\cite{Cirelli:2016rnw}.

\subsection{Vector mediator \label{sec:VectorMed}}

We consider the interaction Lagrangian
\beq
{\cal L} = \bar{X}(i \slashed{D} - \MDM)X 
- \frac{1}{4}{\FD}_{\mu\nu} \FD^{\mu\nu} \,,
\label{eq:L_vector}
\eeq
where $X$ denotes the DM particle, with covariant derivative $D^\mu = \partial^\mu + i g_d \VD^\mu$, and $F_{\mathsmaller{D}}^{\mu\nu} = \partial^\mu \VD^\nu - \partial^\nu \VD^\mu$, with $\VD^\mu$ being the dark photon field and $\aD \equiv g_d^2/(4\pi)$ being the dark fine-structure constant. If $X$ carries a particle-antiparticle asymmetry, another field is required to balance the implied $U(1)_{D}$ charge asymmetry in $X$; we return to the implications of this in section~\ref{Sec:AnnihilationSignals}.

\begin{figure}
\centering
\begin{tikzpicture}[line width=1.1pt, scale=1]
\begin{scope}
\node at (-1,-2) {(a)};
\node at (-1.7, 1.3) {$X$};
\node at (-1.7,-0.3) {$\bar{X}$};
\node at (-2.2,0.45) {$\VD$};
\draw[fermion] (-2,1) -- ( 0,1);
\draw[fermion] ( 0,1) -- ( 0,0);
\draw[fermion] ( 0,0) -- (-2,0);
\draw[vector] (-1.8,0) -- (-1.8,1);
\draw[vector] (-1.4,0) -- (-1.4,1);
\node at (-1,0.5) {$\cdots$};
\draw[vector] (-0.6,0) -- (-0.6,1);
\draw[vector] (0,1) -- (1,1);
\draw[vector] (0,0) -- (1,0);
\node at (1.3,1) {$\VD$};
\node at (1.3,0) {$\VD$};
\end{scope}
%
%
%
%
\begin{scope}[shift={(8,0)}]
\node at (0,-2) {(b)};
\begin{scope}[shift={(-2.5,0)}]
\node at (-1.7, 1.3) {$X$};
\node at (-1.7,-0.3) {$\bar{X}$};
\node at (-2.2,0.45) {$\VD$};
\draw[fermion] (-2,1) -- ( 0,1);
\draw[fermion] ( 0,0) -- (-2,0);
\draw[vector] (-1.8,0) -- (-1.8,1);
\draw[vector] (-1.4,0) -- (-1.4,1);
\node at (-1,0.5) {$\cdots$};
\draw[vector] (-0.6,0) -- (-0.6,1);
\draw[vector] (0,1) -- (0.4,1.7);
\node at (-0.2,1.4) {$\VD$};
\draw[fermion] (0,1) -- (2,1);
\draw[fermion] (2,0) -- (0,0);
\draw[vector] (1.8,0) -- (1.8,1);
\draw[vector] (1.4,0) -- (1.4,1);
\node at      (1,0.5) {$\cdots$};
\draw[vector] (0.6,0) -- (0.6,1);
\draw[thin] (1.2,0.5) ellipse (0.95 and 0.85);
\end{scope}
\begin{scope}[shift={(2.5,1)}]
\draw[fermion] (-2,1) -- ( 0,1);
\draw[fermion] ( 0,1) -- ( 0,0);
\draw[fermion] ( 0,0) -- (-2,0);
\draw[vector] (-1.8,0) -- (-1.8,1);
\draw[vector] (-1.4,0) -- (-1.4,1);
\node at      (-1,0.5) {$\cdots$};
\draw[vector] (-0.6,0) -- (-0.6,1);
\draw[thin] (-1.2,0.5) ellipse (0.95 and 0.85);
\draw[vector] (0,1) -- (1,1);
\draw[vector] (0,0) -- (1,0);
\node at (1.3,1) {$\VD$};
\node at (1.3,0) {$\VD$};
\end{scope}
\begin{scope}[shift={(2.5,-1)}]
\draw[fermion] (-2,1)   -- ( 0,1);
\draw[fermion] ( 0,1)   -- ( 0,0.5);
\draw[fermion] ( 0,0.5) -- ( 0,0);
\draw[fermion] ( 0,0)   -- (-2,0);
\draw[vector] (-1.8,0) -- (-1.8,1);
\draw[vector] (-1.4,0) -- (-1.4,1);
\node at      (-1,0.5) {$\cdots$};
\draw[vector] (-0.6,0) -- (-0.6,1);
\draw[thin] (-1.2,0.5) ellipse (0.95 and 0.85);
\draw[vector] (0,1)   -- (1,1);
\draw[vector] (0,0.5) -- (1,0.5);
\draw[vector] (0,0)   -- (1,0);
\node at (1.3,1)   {$\VD$};
\node at (1.3,0.5) {$\VD$};
\node at (1.3,0)   {$\VD$};
\end{scope}
\end{scope}
\end{tikzpicture}
\caption[Feynman diagrams for annihilation and bound-state formation and decay]
{\label{fig:FeynmanDiagrams}
Dark matter coupled to a light or massless dark photon $\VD$ can annihilate either 
(a) directly into radiation, or 
(b) in two steps, via the radiative formation of particle-antiparticle bound states, and their subsequent decay into two or three dark photons, for the spin-singlet (para) and spin-triplet (ortho) configurations respectively. 
Both the direct annihilation and the formation of bound states are enhanced by the Sommerfeld effect (initial-state ladder); in the Coulomb regime, bound-state formation is faster than annihilation whenever the Sommerfeld effect is important ($v_{\rm rel} \lesssim \alpha_D$).
}
\end{figure}
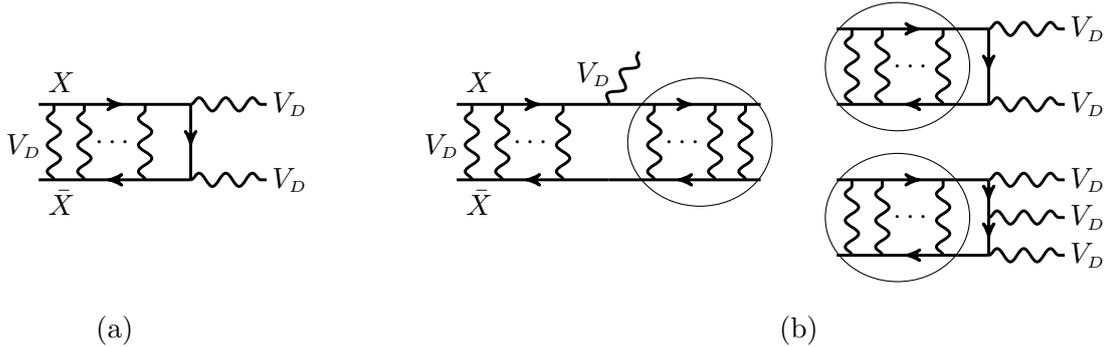

Two processes contribute significantly to the depletion of DM in the early universe~\cite{vonHarling:2014kha}: the direct annihilation into two dark photons, and the radiative formation of positronium-like bound states followed by their decay, 
\begin{align}
X+\bar{X} \to\ & 2\VD \,,
\label{eq:ANN}
\\
X+\bar{X} \to\ & {\cal B}_s(\bar{X}X) + \VD(\omega) \,,
\label{eq:BSF}
\\
&{\cal B}_{\uparrow\downarrow}(\bar{X}X) \to 2\VD \,,
\tag{\ref{eq:BSF}a}
\label{eq:DecayPara}
\\
&{\cal B}_{\uparrow\uparrow}(\bar{X}X) \to 3\VD \,,
\tag{\ref{eq:BSF}b}
\label{eq:DecayOrtho}
\end{align}  
where the subscript $s = \, \uparrow\downarrow, \, \uparrow\uparrow$ denotes the spin-singlet and spin-triplet bound states, which form with 25\% and 75\% probability, respectively. The dark photon emitted during bound-state formation (BSF) carries away energy $\omega = \Delta + E_k$, where $\Delta = \MDM \aD^2/4$ is the binding energy and $E_k = \MDM \vrel^2/4$ is the kinetic energy of the two incoming particles in the center-of-momentum frame. The Feynman diagrams for the processes \eqref{eq:ANN} and \eqref{eq:BSF} are shown in \cref{fig:FeynmanDiagrams}.

The (spin-averaged) cross-sections for annihilation and radiative capture to the ground state can be expressed as~\cite{vonHarling:2014kha,Petraki:2015hla,Petraki:2016cnz}
\begin{subequations}
\label{eq:VectorMed_sigmas}
\label[pluralequation]{eqs:VectorMed_sigmas}
\begin{align}
\sigma_\ann \vrel &= \sigma_0 \, S_\ann^{(0)} \,,
\label{eq:VectorMed_sigma_ann}
\\
\sigma_\BSF \vrel &= \sigma_0 \, S_\BSF  \,,
\label{eq:VectorMed_sigma_BSF}
\end{align}
\end{subequations}
where 
\beq \sigma_0 \equiv \pi \aD^2/\MDM^2 \label{eq:sigma0} \eeq 
is the perturbative value of the annihilation cross-section times relative velocity.
$S_\ann^{(0)}$ is the Sommerfeld enhancement factor for the $s$-wave annihilation process \eqref{eq:ANN}. In contrast, the capture with emission of a vector boson \eqref{eq:BSF} is a $p$-wave process, ${\cal M}_\BSF \propto \sin \theta \propto d_{1,0}^1(\theta)$.\footnote{%
Here, $d_{\lambda_f, \lambda_i}^J(\theta)$ are the Wigner $d$ functions. $\theta$ is the scattering angle, $J$ denotes the partial wave, and $\l_i = \l_{i1} - \l_{i2}$, $\l_f = \l_{f1} - \l_{f2}$ are the initial- and final-state helicities respectively, with the indices 1 and 2 denoting the two particles of each state. Note that the partial-wave decomposition of ${\cal M}_\BSF$ was not correctly described in Refs.~\cite{vonHarling:2014kha, Petraki:2015hla}.} 

In the Coulomb limit, $S_\ann^{(0)}$ and $S_\BSF$ depend only on the ratio $\z \equiv \aD/\vrel$, and can be computed analytically~\cite{vonHarling:2014kha, Petraki:2015hla, Petraki:2016cnz}, 
\begin{subequations}
\label{eq:VectorMed_Sfactors}
\label[pluralequation]{eqs:Sfactors_VectorMed}
\begin{align}
S_\ann^{(0)}(\zeta) &= \frac{2\pi \z}{1-e^{-2\pi \z}} \,,
\label{eq:S_ann_0}
\\
S_\BSF(\zeta) &= \frac{2\pi \z}{1-e^{-2\pi \z}} 
\ \frac{\z^4}{(1+\z^2)^2} 
\ \frac{2^9}{3} \ e^{-4\z \: {\rm arccot} (\z)} \,.
\label{eq:S_BSF}
\end{align}
\end{subequations}
In this parametrisation, it is easily seen that in the regime where the Sommerfeld effect is important, $\vrel \lesssim \aD$, both the annihilation and BSF cross-sections exhibit the same velocity dependence, $\sigma \vrel \propto 1/\vrel$, with BSF being the dominant inelastic process, $\sigma_\BSF / \sigma_\ann \simeq 3.13$~\cite{vonHarling:2014kha}.\footnote{
In the same regime, the capture into $n=2, \ell=1$ bound states is also somewhat faster than annihilation~\cite{Petraki:2016cnz}. However, it is subdominant with respect to the capture to the ground state ($n=1, \ell=0$), and has a smaller decay rate, which renders it less efficient in depleting DM in the early universe. We shall ignore it in our analysis.}
On the other hand, for $\vrel > \aD$, BSF is very suppressed and subdominant to annihilation.
The bound-state decay rates are $\Gamma_{\rm dec, \uparrow\downarrow} = \aD^5 \MDM/2$ and $\Gamma_{\rm dec, \uparrow\uparrow} = c_{\aD} \Gamma_{\rm dec, \uparrow\downarrow}$, where $c_{\aD} \equiv 4(\pi^2-9)\aD/(9\pi)$.

\begin{figure}[t!]
\centering
{\bf Vector mediator} 
\smallskip
\includegraphics[height=5cm]{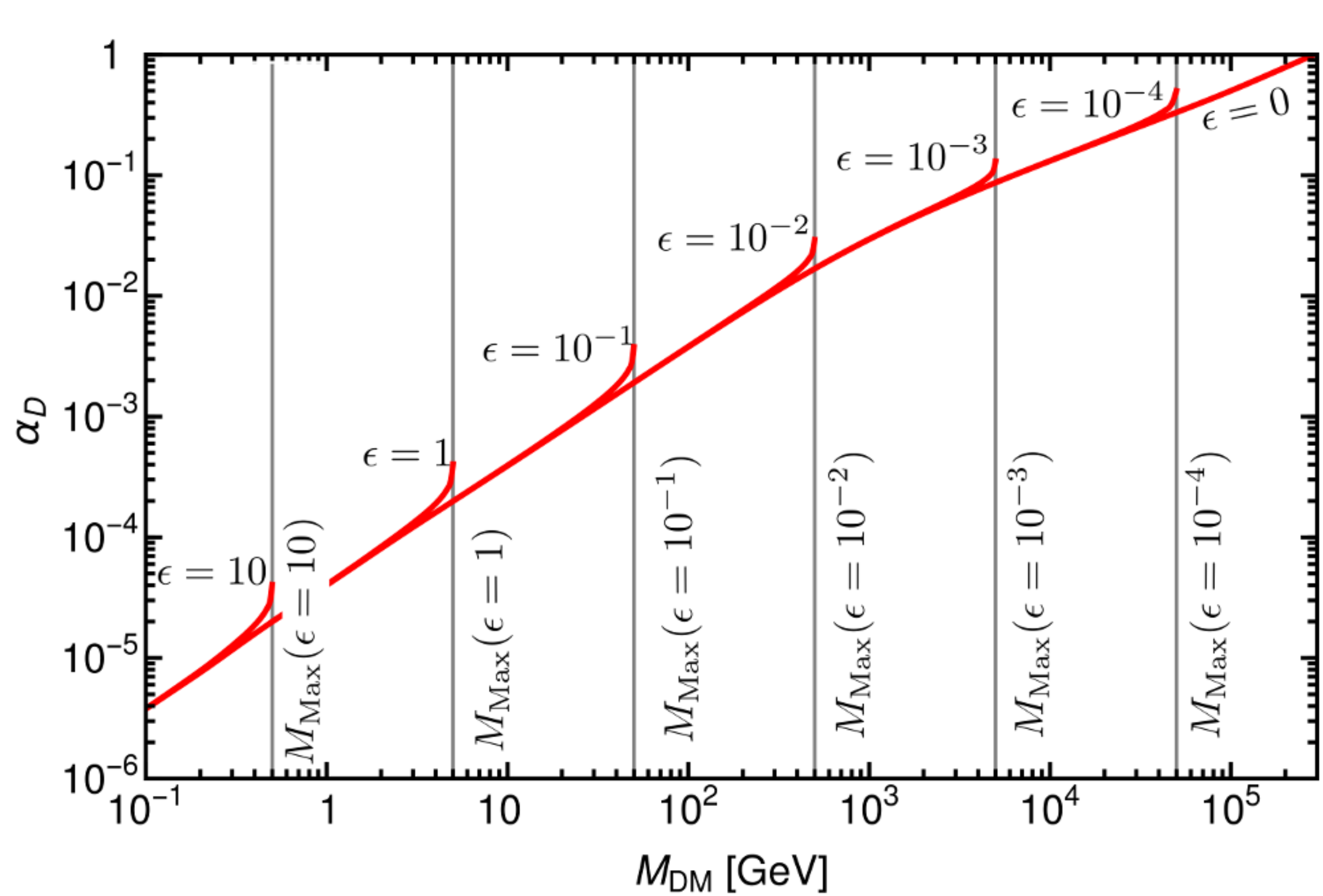}~~~
\includegraphics[height=5cm]{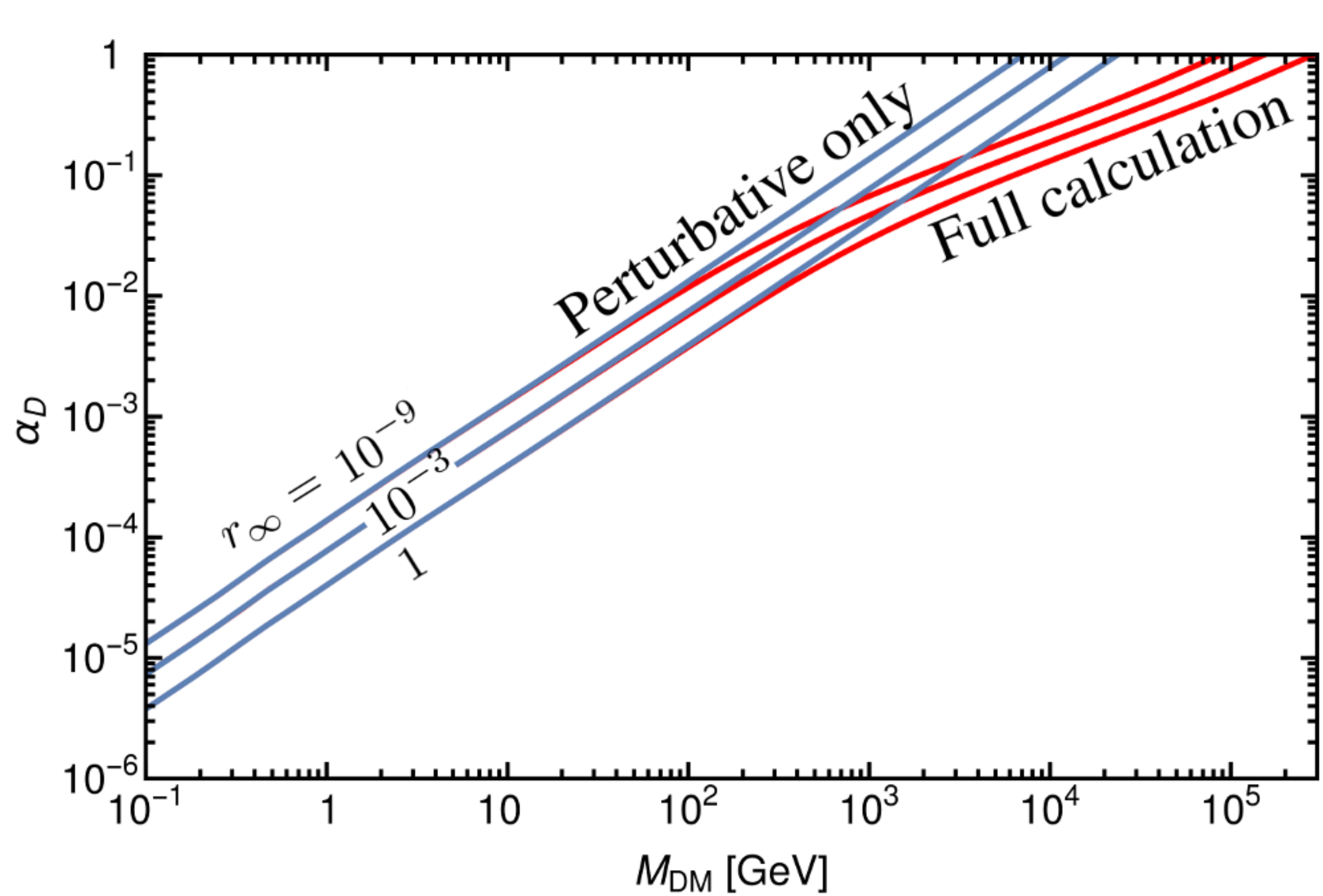}

\includegraphics[height=5cm]{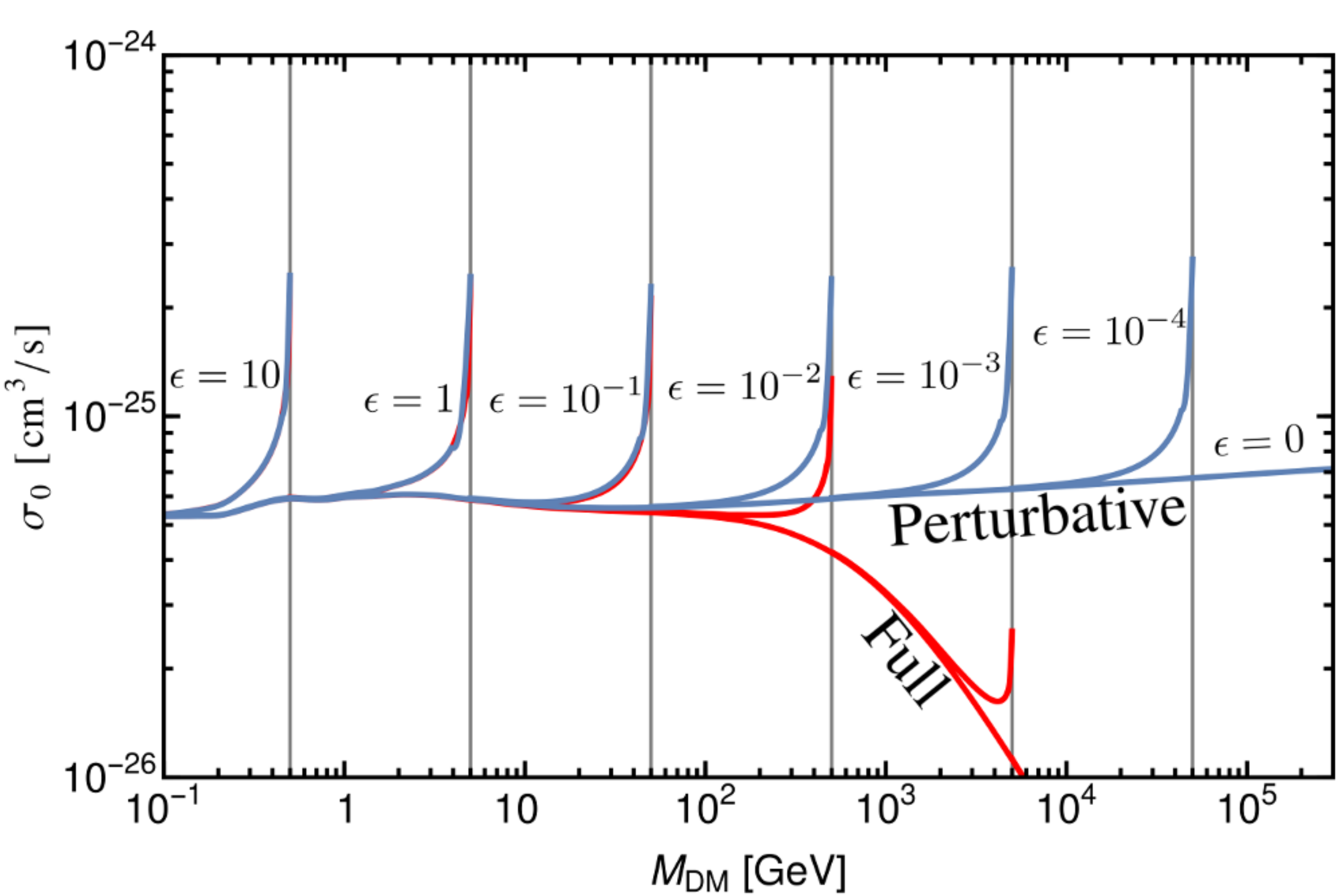}~~~
\includegraphics[height=5cm]{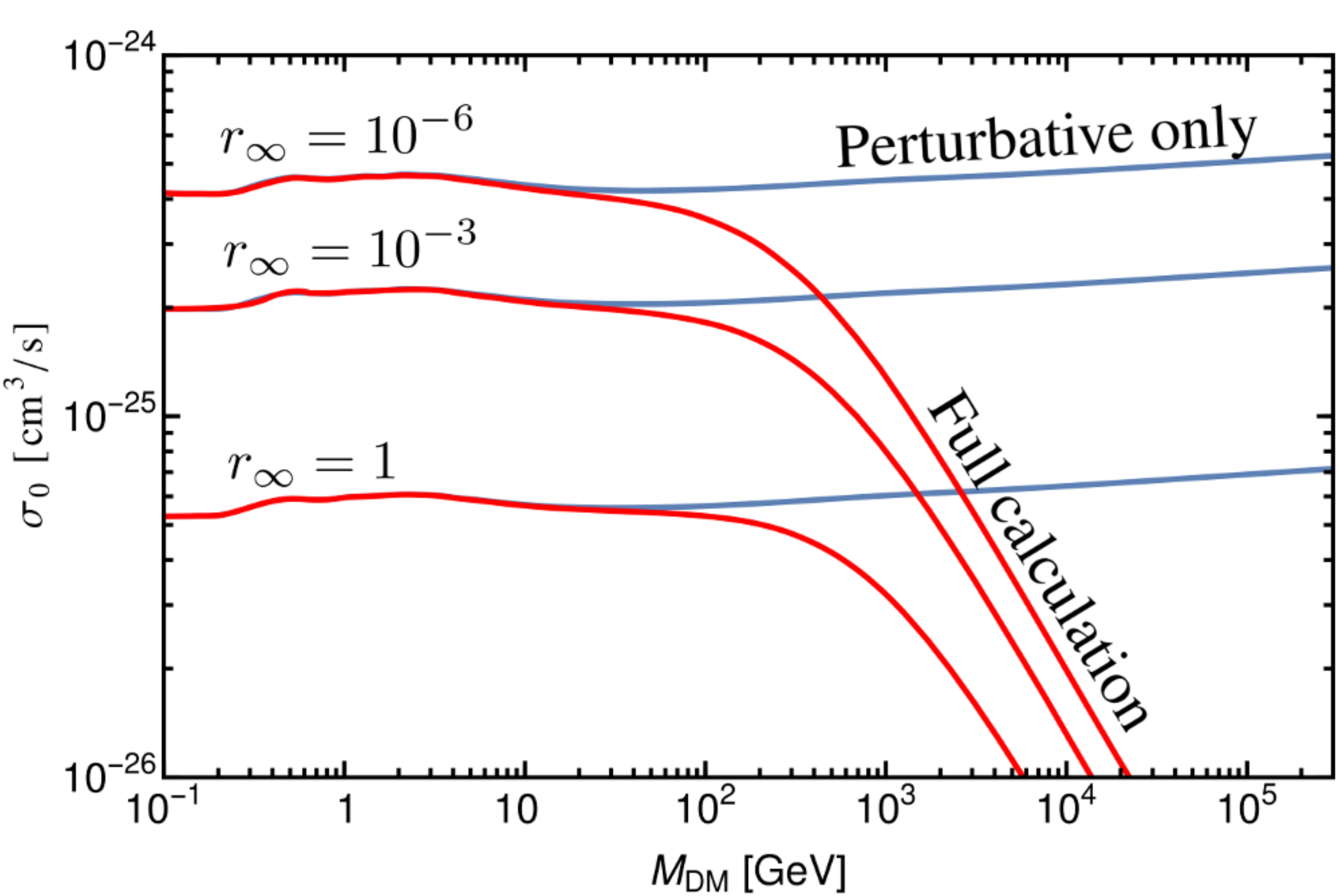}

\caption[]{\label{fig:VectorMed_coupling}
\emph{Top:} The dark fine-structure constant required to establish the observed DM density via thermal freeze-out, $\alpha_D^{}$ vs $M_{\rm DM}$, for fixed values of the dark particle-minus-antiparticle-number-to-entropy ratio $\eta_D^{} = \epsilon \eta_B^{}$ (\emph{left}), and for fixed values of the final antiparticle-to-particle ratio $r_\infty \equiv (n^+/n^-)_{t\to\infty}$ (\emph{right}). The red lines include both the Sommerfeld enhancement of the direct DM annihilation into radiation, and the formation and decay of particle-antiparticle bound states. The blue lines ignore all non-perturbative effects.

\smallskip
\emph{Bottom:} The perturbative annihilation cross-section times relative velocity, $\sigma_0 = \pi \alpha_D^2/M_{\rm DM}^2$ vs $M_{\rm DM}$, for fixed $\epsilon = \eta_D^{} / \eta_B^{}$ (\emph{left}), and $\rinf$ (\emph{right}). 
$\sigma_0$ is evaluated using $\alpha_D^{}$ determined as described above.

\smallskip
For $M_{\rm DM} \ll M_{\rm max} (\epsilon)$ or $r_\infty \approx 1$, $\alpha_D$ and $\sigma_0$ closely track the symmetric DM curve ($\epsilon=0$). For $r_\infty \ll 1$ to be attained and $M_{\rm DM} \simeq M_{\rm max}$ to be realised, a stronger coupling is required.
The stronger coupling implies that the Sommerfeld effect --- which reduces the expected coupling in comparison to perturbative annihilation --- is more pronounced for smaller $r_\infty$, and extends to lower $M_{\rm DM}$ values.}

\end{figure}

The evolution of the DM density in the early universe is governed by a set of coupled equations that tracks the densities of the unbound DM particles and anti-particles, as well as the densities of the bound states. These equations capture the effect of direct DM annihilation and pair creation, as well as the interplay between bound-state formation, ionisation and decay processes that determines the efficiency of BSF in depleting DM. Because the velocity dependence of $\sigma_\ann$ and $\sigma_\BSF$ arises via the parameter $\zeta = \aD/\vrel= (\Delta/E_k)^{1/2}$, the thermally-averaged cross-sections depend on 
\beq \zD \equiv \Delta/\TD = (\aD^2/4)\,\xD \,. \label{eq:zD} \eeq
We shall use $\zD$ as the time variable, instead of $\xD$, and denote the number-density-to-entropy ratios for  the spin-singlet and triplet states with $Y_{\uparrow\downarrow}$ and $Y_{\uparrow\uparrow}$ respectively. Adapting the Boltzmann equations from Ref.~\cite{vonHarling:2014kha} to accommodate for a non-zero particle-antiparticle asymmetry, we obtain
\begin{subequations}
\label{eq:VectorMed_Boltz}
\label[pluralequation]{eqs:VectorMed_Boltz}
\begin{align}
\frac{dr}{d\zD} = \:
& - \frac{\etaD \lambda_1 \, g_*^{1/2} \bar{S}_\ann^{(0)} (\zD)}{\zD^2} 
\: \[r - r_{\rm eq} \(\frac{1-r}{1-r_{\rm eq}}\)^2\]
\nn \\
& - \frac{\etaD \lambda_1 \, g_*^{1/2} \bar{S}_\BSF (\zD) }{\zD^2} \: r   
  + \frac{\lambda_2  \, g_*^{1/2} \, \zD \, f_{\rm ion}(\zD) \, ( Y_{\uparrow\downarrow} + Y_{\uparrow\uparrow}) (1-r)^2}{\etaD \, h_{\rm eff} }   \,, 
\label{eq:VectorMed_dr/dz}  \\
\frac{d Y_{\uparrow\downarrow}}{d\zD} = \:
& \frac{\etaD^2 \lambda_1 \, g_*^{1/2} \bar{S}_\BSF (\zD)}{4\zD^2}  \: \frac{r}{(1-r)^2}
- \frac{\lambda_2 g_*^{1/2}  \zD f_{\rm ion}(\zD) \, Y_{\uparrow\downarrow}}{h_{\rm eff} } 
- \frac{\lambda_2  g_*^{1/2}  \zD \, (Y_{\uparrow\downarrow} - Y_{\uparrow\downarrow}^{\rm eq})}{h_{\rm eff}} 
\,,
\label{eq:VectorMed_dpara/dz}  \\
\frac{d Y_{\uparrow\uparrow}}{d\zD} = \:
& \frac{3 \etaD^2 \lambda_1 \, g_*^{1/2} \bar{S}_\BSF (\zD)}{4\zD^2}  \: \frac{r}{(1-r)^2}
- \frac{ \lambda_2 g_*^{1/2}  \zD f_{\rm ion}(\zD) \, Y_{\uparrow\uparrow}}{h_{\rm eff}} 
- \frac{\lambda_2 g_*^{1/2}  c_{\aD}  \, \zD \, (Y_{\uparrow\uparrow} - Y_{\uparrow\uparrow}^{\rm eq})}{h_{\rm eff}} 
\,,
\label{eq:VectorMed_dortho/dz}
\end{align}
\end{subequations}
where $\lambda_1 \equiv \sqrt{\pi/45}\, \sigma_0 \Delta\MPl$, 
$\lambda_2 \equiv \sqrt{45/(4\pi^3)} \, (\aD^5\MDM/2) \, (\MPl/\Delta^2)$ and
\begin{subequations}
\begin{align}
\bar{S}_\ann^{(0)} (\zD) &\equiv \frac{2}{\sqrt{\pi}}
\int_0^\infty du \ S_\ann^{(0)} (\sqrt{\zD/u}) \ \sqrt{u} \, \exp(-u) \,,
\label{eq:Sbarann}
\\
\bar{S}_\BSF (\zD) &\equiv \frac{2}{\sqrt{\pi}}
\int_0^\infty du \ S_\BSF (\sqrt{\zD/u}) 
\ \frac{\sqrt{u} \ \exp(-u)}{1- \exp(-\zD-u)} \,,
\label{eq:SbarBSF}
\\
f_{\rm ion}(\zD) &\equiv 
 \frac{ \Gamma_{\rm ion}(\zD) }{ \Gamma_{\rm dec, \uparrow\downarrow} }
=\frac{(1-\aD^2/8)^{-3/2}}{8\pi} \int_0^\infty \frac{d\zeta}{\zeta^4} 
\ \frac{S_\BSF(\zeta)}{\exp\[\zD(1+1/\zeta^2)\]-1} \,.
\label{eq:fion}
\end{align}
\end{subequations}
Here, $\bar{S}_\ann^{(0)}$ is the thermal average of $S_\ann^{(0)}$, while $\bar{S}_\BSF$ is the thermally averaged  $S_\BSF$ times the Bose enhancement due to the final-state dark photon emitted during BSF [cf.~\cref{eq:BSF}]. $\Gamma_{\rm ion}(\zD)$ is the bound-state ionisation rate, averaged over the dark photon thermal bath, and depends on $S_\BSF$ because the rates of inverse processes are related via detailed balance. (Ensuring that detailed balance is maintained even at large $\aD$ is the reason we have included the  $(1-\aD^2/8)^{-3/2}$ factor in \cref{eq:fion}, despite it going beyond the order of the approximation used in computing $S_\BSF$.)  \Cref{eqs:VectorMed_Boltz} include also the bound-state inverse decays, which were omitted in Ref.~\cite{vonHarling:2014kha}. Their effect is negligible. We solve the system of equations~(\ref{eq:VectorMed_dortho/dz}) and deduce the required coupling to obtain the observed DM density. We present our results in \cref{fig:VectorMed_coupling}.

\begin{figure}[t!]
\centering
{\bf Vector mediator} \\
\includegraphics[height=5cm]{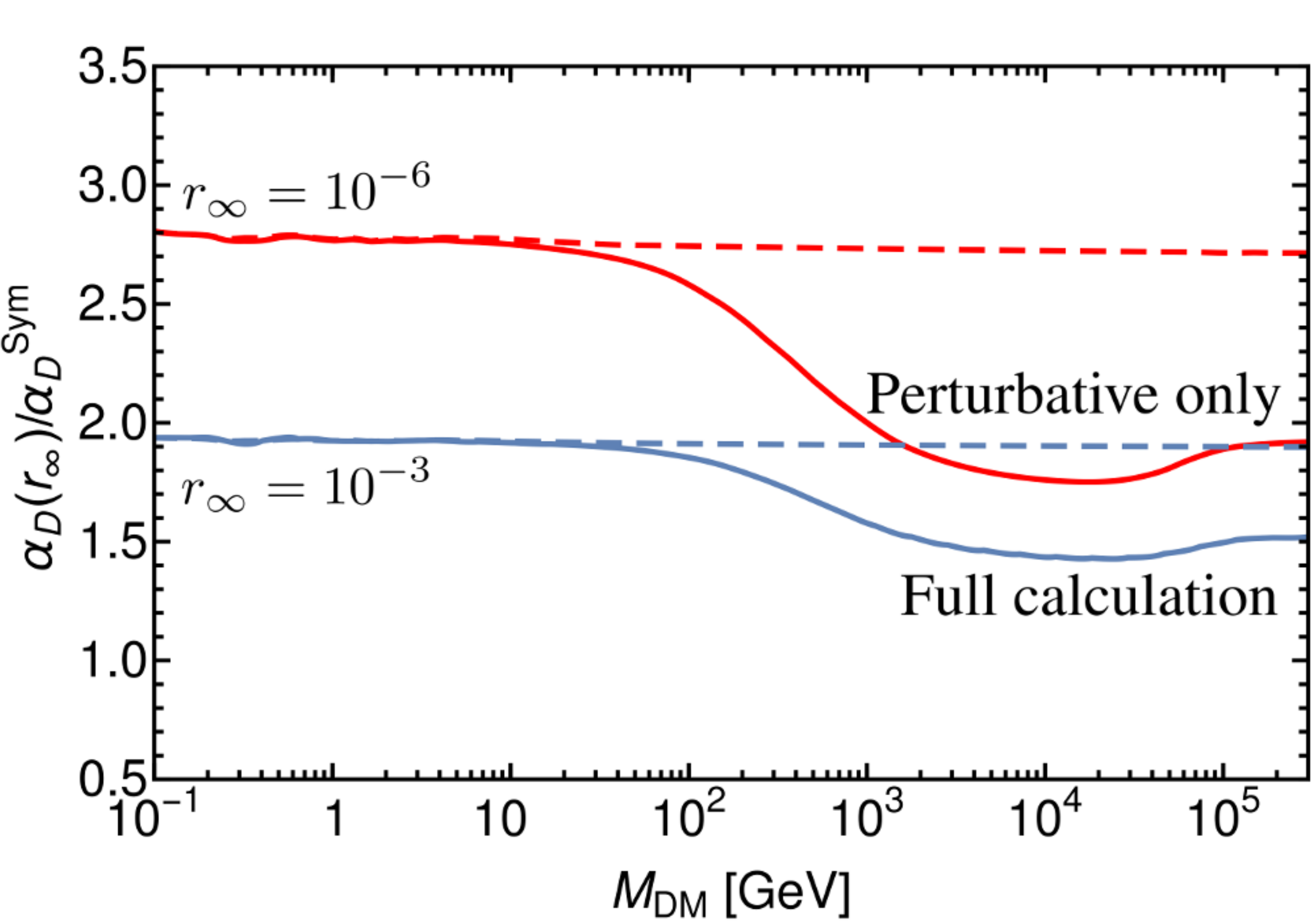}~~~
\includegraphics[height=5cm]{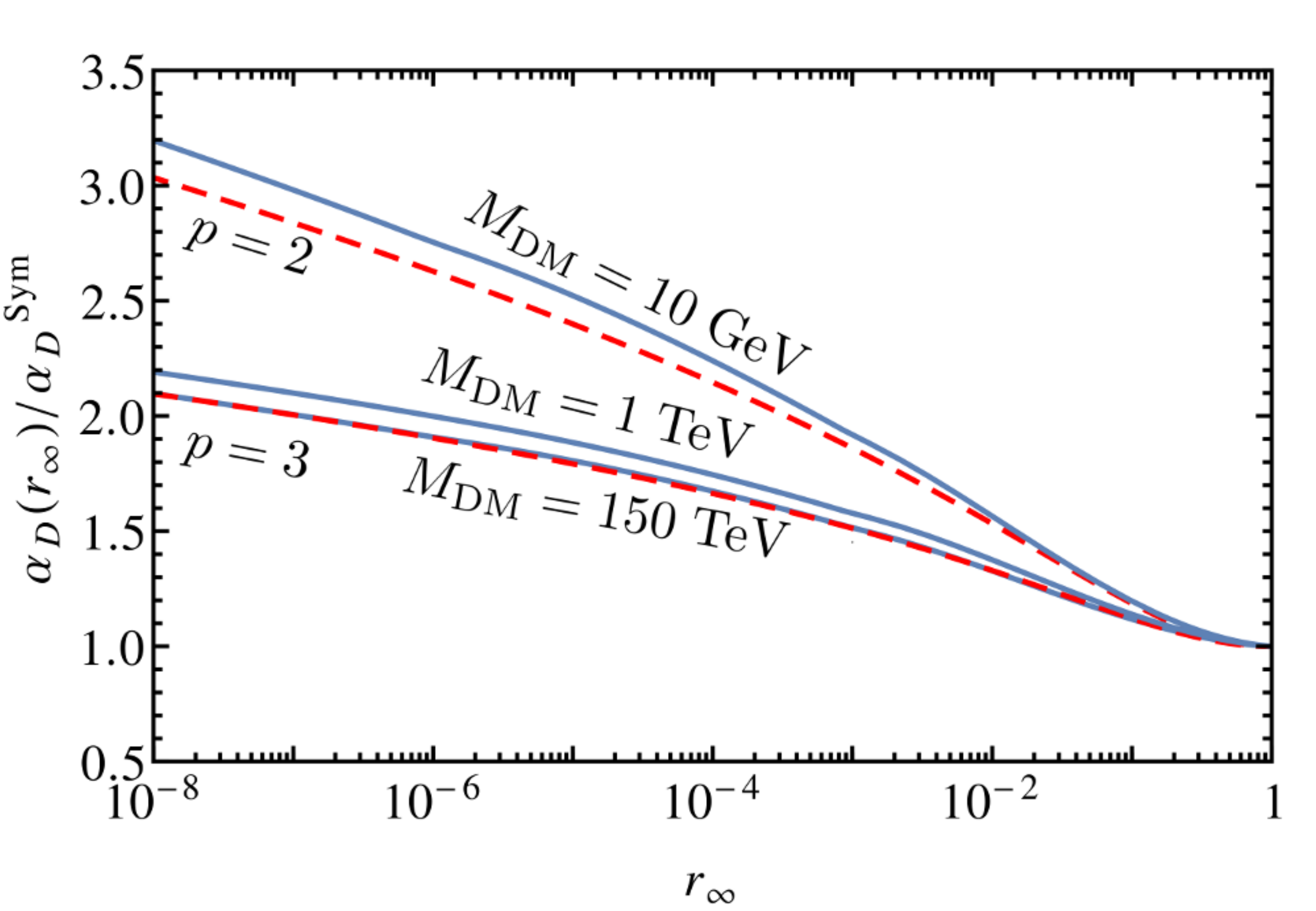}
\caption[]{\label{fig:VectorMed_AlphaRatios} 
The ratio of the coupling required to establish a fractional asymmetry $r_\infty$ to the coupling required for symmetric DM of the same mass. The larger the asymmetry (i.e. the smaller $r_\infty$), the larger $\alpha_D^{}(r_\infty)/\alpha_D^\sym$.

\smallskip
\emph{Left:}  $\alpha_D^{}(r_\infty)/\alpha_D^\sym$ vs $M_{\rm DM}$, for $r_{\infty} = 10^{-3}$ (blue) and $r_{\infty}=10^{-6}$ (red). 
Annihilation via Sommerfeld enhancement processes (solid lines) implies a lower $\alpha_D^{}(r_\infty)/\alpha_D^\sym$
in comparison to perturbative annihilation (dashed lines).
For small (large) $M_{\rm DM}$, freeze-out occurs well within the perturbative (Sommerfeld enhanced) regime, where the inelastic cross-sections scale as $\sigma_\inel^\FO\propto \alpha_D^p$, with $p=2$ ($p=3$). Within these regimes, $\alpha_D^{}(r_\infty)/\alpha_D^\sym$ is largely independent of the DM mass for a fixed $r_\infty$, as anticipated by the analytical approximation of eq.~\eqref{eq:AlphaRatio}. (The mild dependence of $\alpha_D^{}(r_\infty)/\alpha_D^\sym$ on $M_{\rm DM}$ at large $M_{\rm DM}$, is due to the intricacy of the effect of bound states on the DM relic density. See text for discussion.)

\smallskip
\emph{Right:} $\alpha_D^{}(r_\infty)/\alpha_D^\sym$ vs $r_\infty$, for $M_{\rm DM} = 10$~GeV, 1~TeV and 150~TeV (blue solid lines, from top to bottom).  The red dashed lines denote the analytical approximation of eqs.~\eqref{eq:rinf_final_full} and \eqref{eq:AlphaRatio}, assuming that the inelastic cross-sections around the time of freeze-out scale as $\sigma_\inel^\FO\propto \alpha_D^p$, with $p=2$ for fully perturbative annihilation (upper line), and $p=3$ for Sommerfeld-enhanced annihilation (lower line). For intermediate masses (e.g.~$M_{\rm DM} \sim 1$~TeV), the scaling of $\sigma_\inel^\FO$ with $\alpha_D^{}$, and consequently the scaling of $\alpha_D^{}(r_\infty)/\alpha_D^\sym$ with $r_\infty$, fall in between these two cases.}
\end{figure}

\medskip
As described in Ref.~\cite{vonHarling:2014kha}, the formation of bound states depletes efficiently the DM density only after the bound-state decay becomes faster than ionisation. The decay rate of the spin-singlet bound state becomes larger than ionization at $f_{\rm ion}(\zD) \lesssim 1$, or $\zD \gtrsim 0.28$. For the spin-triplet state, this occurs at a later time, when $f_{\rm ion}(\zD) \lesssim c_{\aD}$. We may thus adopt the following approximation for the evolution of $r$,
\begin{subequations}
\label{eq:VectorMed_EffectiveMethod}
\label[pluralequation]{eqs:VectorMed_EffectiveMethod}
\beq
\frac{dr}{d\zD} = 
- \frac{\etaD \lambda_1 \, g_*^{1/2} S_{\rm eff}(\zD)}{\zD^2} 
\: \[r - r_{\rm eq} \(\frac{1-r}{1-r_{\rm eq}}\)^2\] \,,
\tag{\ref{eq:VectorMed_EffectiveMethod}}
\label{eq:VectorMed_dr/dz_eff}
\eeq
where $S_{\rm eff}$ is defined as~\cite{vonHarling:2014kha}
\begin{equation}    
S_{\rm eff}(\zD) \equiv \left\{
\bal{6}   
&\bar{S}_\ann(\zD),	& \qquad & \zD \lesssim 0.28 \,,  \\  
&\bar{S}_\ann(\zD)+\bar{S}_\BSF(\zD)/4,	&\qquad& 0.28 \lesssim  \zD \ {\rm and} \ c_{\aD} \lesssim f_{\rm ion}(\zD)\,, \\  
&\bar{S}_\ann(\zD)+\bar{S}_\BSF(\zD),	&\qquad& f_{\rm ion} (\zD)  \lesssim c_{\aD} \,.  
\eal  
\right.  
\label{eq:Seff_step}  
\end{equation} 
\end{subequations}
This approximation\footnote{%
A similar prescription for $S_{\rm eff}$, with a smoother transition between regimes, has been offered in Ref.~\cite{Liew:2016hqo}. Adapted to the present model, it reads
\beq
S_{\rm eff}(\zD) \equiv \bar{S}_\ann(\zD) +
\frac{1}{1+f_{\rm ion}(\zD)} \, \frac{\bar{S}_\BSF(\zD)}{4} +
\frac{c_{\aD}}{c_{\aD} + f_{\rm ion}(\zD)} \, \frac{3\bar{S}_\BSF(\zD)}{4} \,.
\nn 
\eeq}
produces results that are in very good agreement with those obtained from the full treatment of \cref{eqs:VectorMed_Boltz}. Moreover, \cref{eq:VectorMed_dr/dz_eff} can be mapped to the discussion of \cref{Sec:AsymFO}, and in particular \cref{eq:dr/dz}, by identifying 
\begin{subequations}
\begin{align}
\sigma_* &\to \sigma_0 \,,
\\
F(\xD) &\to S_{\rm eff}(\zD) \,, 
\end{align}
\end{subequations}
where $\zD$ and $\xD$ are related via \cref{eq:zD}.

We discern two regimes, the perturbative and the Sommerfeld-enhanced. The Sommerfeld enhancement is important for $\sqrt{\zD} \simeq \<\a/\vrel\> \gtrsim 1$. 
While $\xD^\FO \approx 25-30$ is insensitive to $\MDM$, $\zD^\FO = (\aD^2/4)\xD^\FO$ increases with $\aD$ and consequently $\MDM$. 
For small $\MDM$, roughly $\MDM \lesssim 100~\GeV$, freeze-out occurs mostly in the perturbative regime, $\zD^\FO \ll 1$; then $F(\xD^\FO) \simeq 1$, $\Phi \simeq \Phi_\sym \simeq \sqrt{g_*^\FO }/\xD^\FO$ and $\sigma_* \Phi \propto \aD^2$. 
On the other hand, for larger $\MDM$, roughly $\MDM \gtrsim 1~\TeV$, freeze-out happens close to or within the Sommerfeld-enhanced regime. When well within the Sommerfeld-enhanced regime, $F(\xD) \sim \zD^{1/2}$, $\Phi \sim \aD\sqrt{g_*^\FO }/\xD^{1/2}$
and $\sigma_* \Phi \propto \aD^3$. Then, from \cref{eq:AlphaRatio}, we may estimate the coupling $\aD(\rinf)$ required to establish a fractional asymmetry $\rinf$, by setting $p=2$ for small $\MDM$, and  $p=3$ for large $\MDM$. It is anticipated that in the small and large $\MDM$ limits, the ratio $\aD/\aD^\sym$ scales solely with $\rinf$, and is insensitive to $\MDM$ for fixed $\rinf$. 

The numerical solution for $\aD/\aD^\sym$ is presented in \cref{fig:VectorMed_AlphaRatios}, and indeed exhibits the two asymptotic behaviours described here. The mild sensitivity of $\aD/\aD^\sym$ on $\MDM$ in the Sommerfeld enhanced regime (large $\MDM$) arises from the numerical coefficient in $F$ and consequently $\sigma_* \Phi$, which depends on whether and which BSF channels contribute to the depletion of the DM density, as described by $S_{\rm eff}$ defined in \cref{eq:Seff_step}.  Since for the same $\MDM$, $\aD$ increases with decreasing $\rinf$, the DM depletion via BSF may be more efficient for $\rinf \ll 1$ than for $\rinf=1$. For example, for $M_{\rm DM} \sim 10$~TeV, only the formation of spin-singlet bound states contributes to the depletion of symmetric DM, while both spin-singlet and spin-triplet bound states deplete DM with $\rinf \sim 10^{-6}$. On the other hand, for $M_{\rm DM} \sim 100$~TeV, both BSF channels contribute to $S_{\rm eff}$ for any $\rinf$.  
Note also that because $\aD$ increases with decreasing $\rinf$, the mass scale of the transition between the perturbative and the Sommerfeld-enhanced regimes depends on $\rinf$.

\subsection{Scalar mediator \label{sec:ScalarMed}}

The interaction Lagrangian is
\beq
{\cal L} 
= \bar{X}(i \slashed{\partial} - \MDM)X 
+ \frac{1}{2}\partial_\m \vf \, \partial^\m \vf- \frac{1}{2} m_\vf^2 \vf^2
- g_d \ \vf \bar{X} X \,,
\label{eq:L_scalar}
\eeq 
with $\vf$ being the dark scalar force mediator with mass $m_\vf$, and $\aD \equiv g_d^2/(4\pi)$. As long as $m_\vf \lesssim \aD (\MDM/2)$, the $X-\bar{X}$ interaction manifests as long range. 

For the determination of the DM relic density, which occurs largely in the Coulomb limit~\cite{Cirelli:2016rnw}, only the DM direct annihilation into two scalars, $X + \bar{X} \to 2\vf$, needs be considered, since it is significantly faster than BSF~\cite{An:2016kie}. The annihilation cross-section times relative velocity is
\begin{subequations}
\label{eq:sigma_ann_ScalarMed}
\label[pluralequation]{eqs:sigma_ann_ScalarMed}
\beq 
\sigma_\ann \vrel = \sigma_1 \, \vrel^2 \ S_\ann^{(1)} \,,
\tag{\ref{eq:sigma_ann_ScalarMed}}
\eeq
where 
\begin{align}
\sigma_1 &= \frac{3\pi \aD^2}{8\MDM^2} \,,
\label{eq:sigma1}
\\
S_\ann^{(1)} (\zeta) &= \frac{2\pi \zeta}{1-e^{-2\pi \zeta}} 
\, (1+\zeta^2) \,.
\label{eq:S_ann_1}
\end{align}
\end{subequations}
Here, $S_\ann^{(1)}$ is the Sommerfeld enhancement factor of $p$-wave annihilation processes, and $\zeta \equiv \aD/\vrel$, as before. At $\vrel \lesssim \aD$, the cross-section exhibits the familiar velocity scaling of Sommerfeld enhanced  processes, $\sigma_\ann \vrel \propto 1/\vrel$. In this regime, the $\vrel^2$ suppression of the perturbative cross-section morphs into an $\aD^2$ suppression, with the entire cross-section scaling as $\sigma_\ann \vrel \propto \aD^5$.

\begin{figure}[t!]
\centering
{\bf Scalar mediator} \\
\includegraphics[height=5cm]{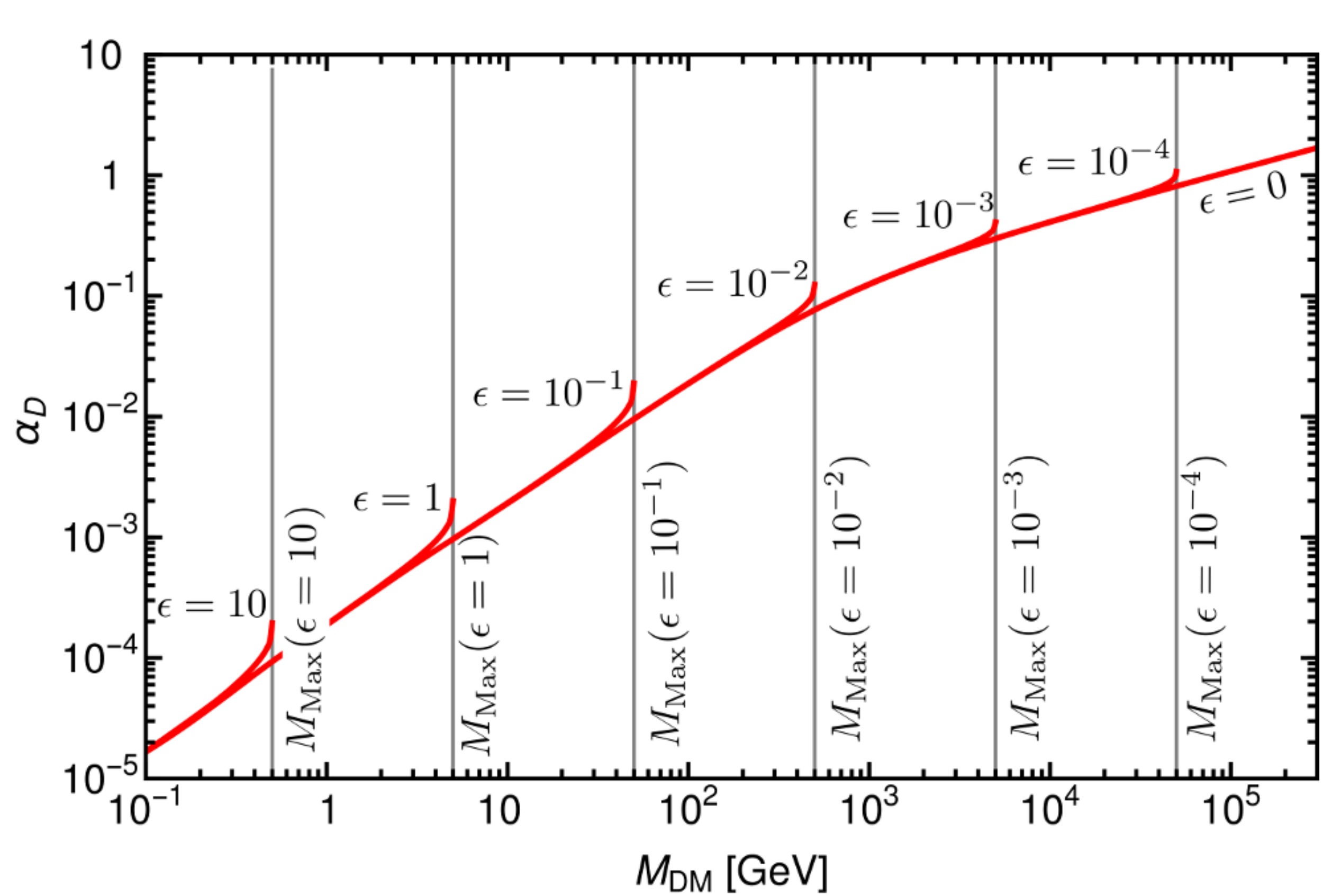}~~
\includegraphics[height=5cm]{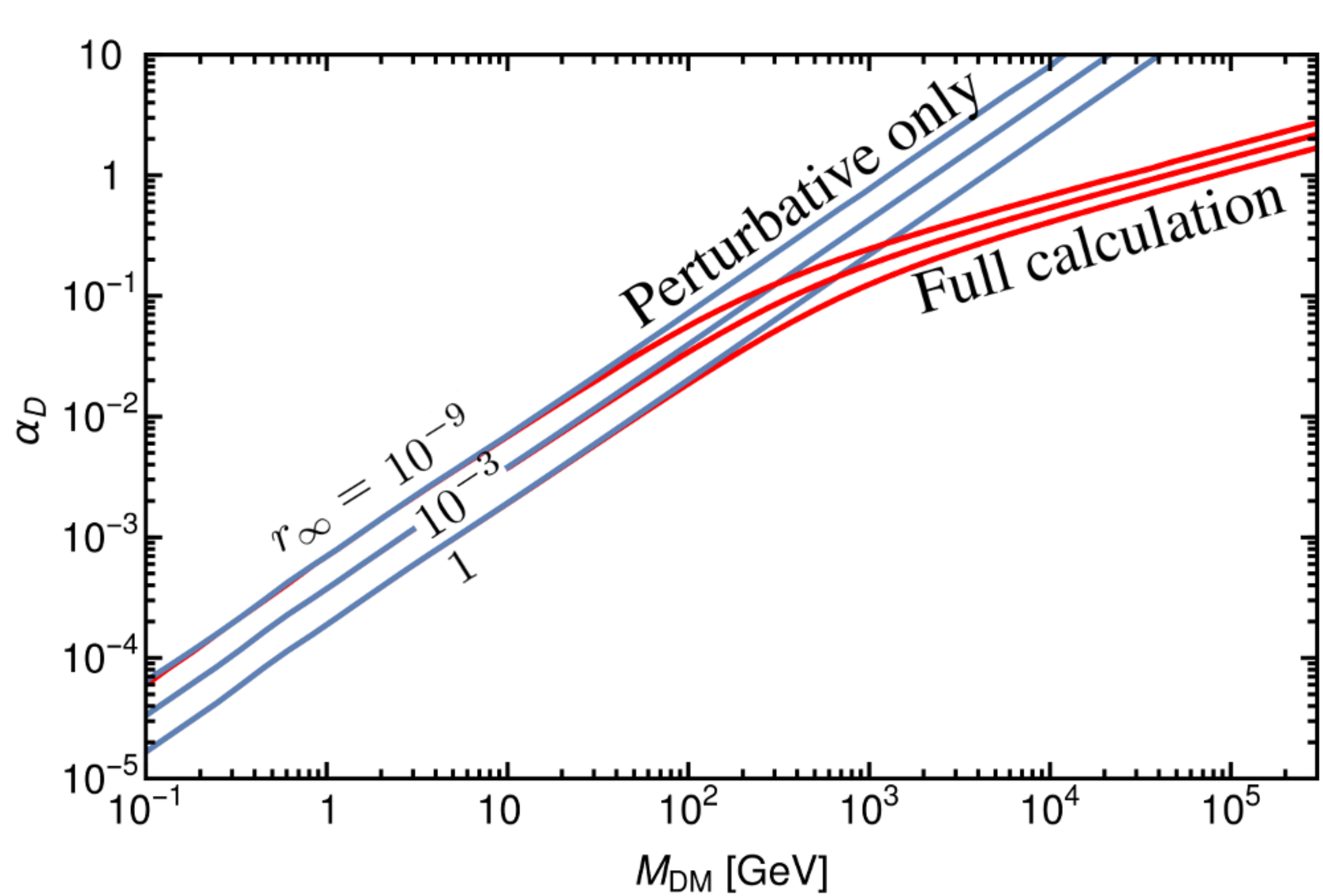}
\\
\includegraphics[height=5cm]{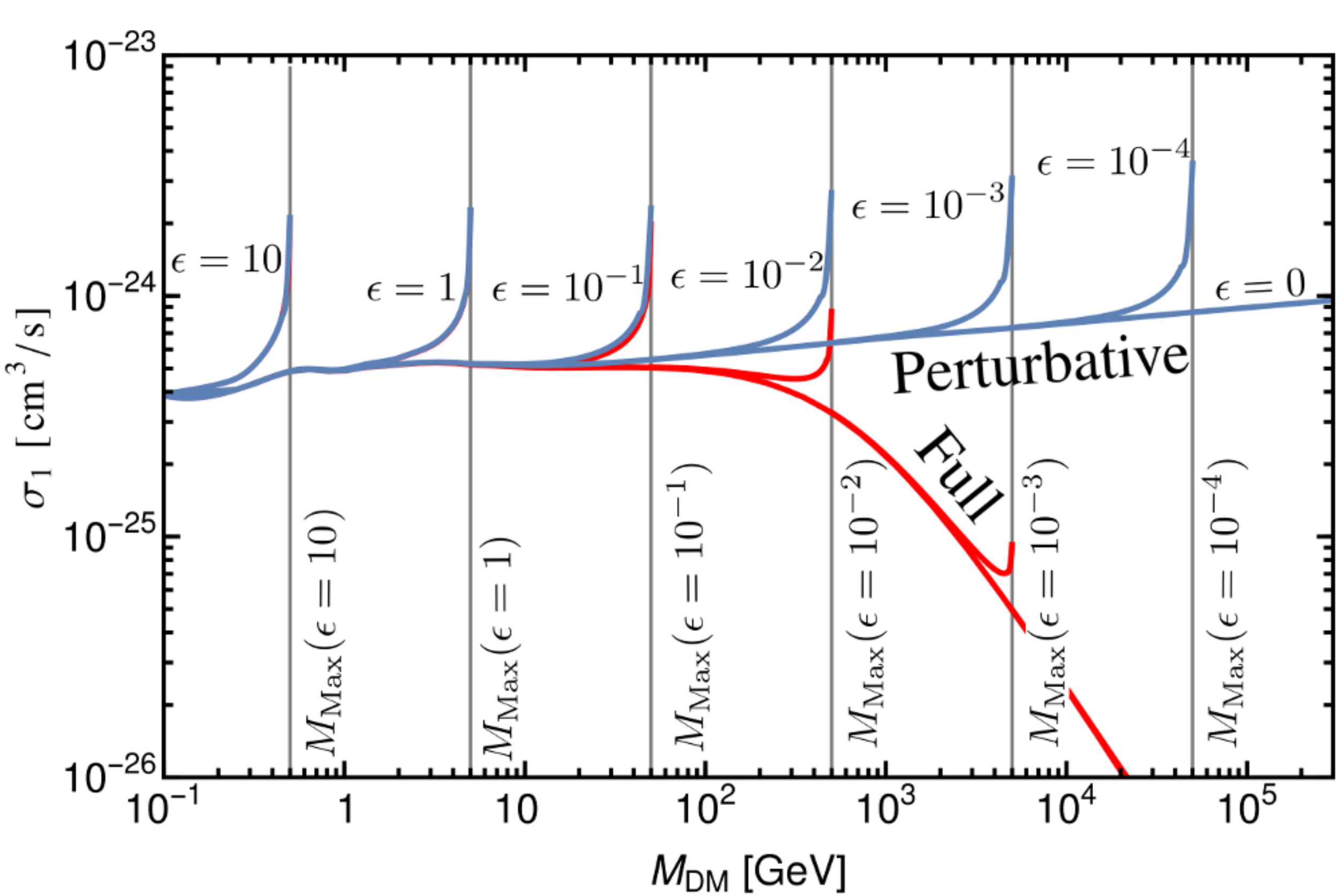}~~
\includegraphics[height=5cm]{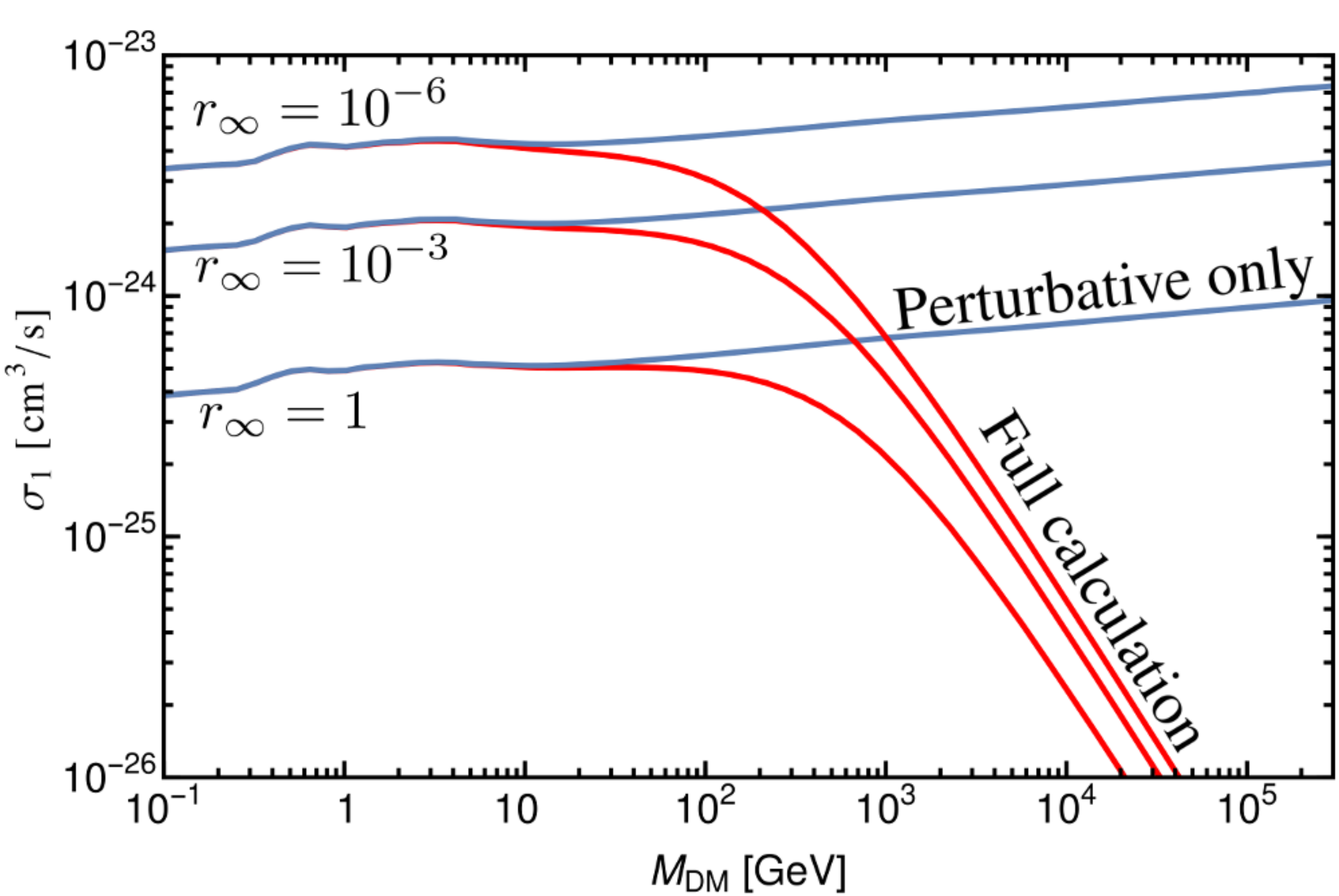}

\caption[]{\label{fig:ScalarMed_coupling}  
The equivalent of fig.~\ref{fig:VectorMed_coupling}, for DM coupled to a light scalar. 
\emph{Top:} $\alpha_D$ vs $M_{\rm DM}$. 
\emph{Bottom:} $\sigma_1 \equiv 3\pi \alpha_D^2/(8M_{\rm DM}^2)$ vs $M_{\rm DM}$.}
\smallskip
\includegraphics[height=5cm]{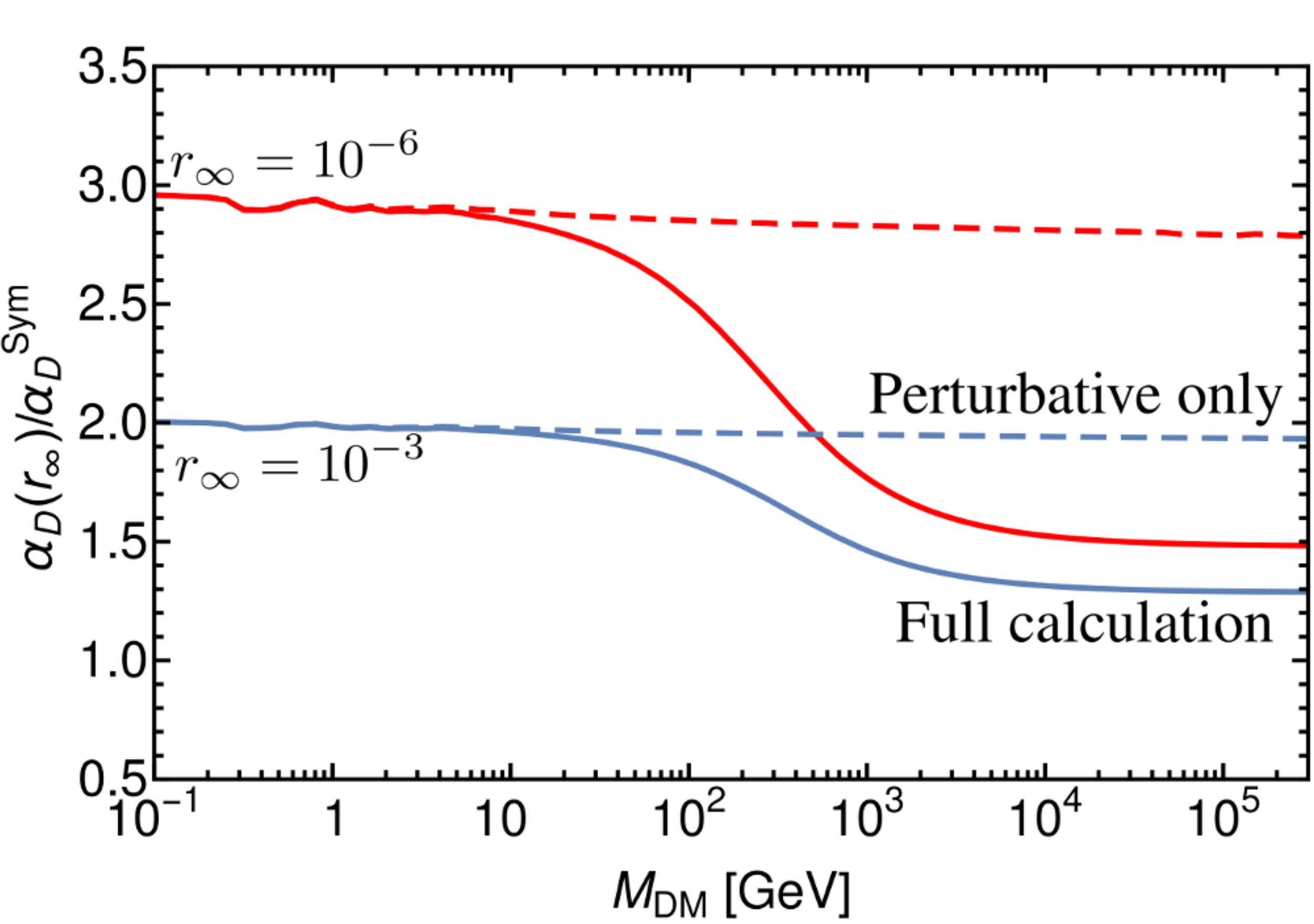}~~~
\includegraphics[height=5cm]{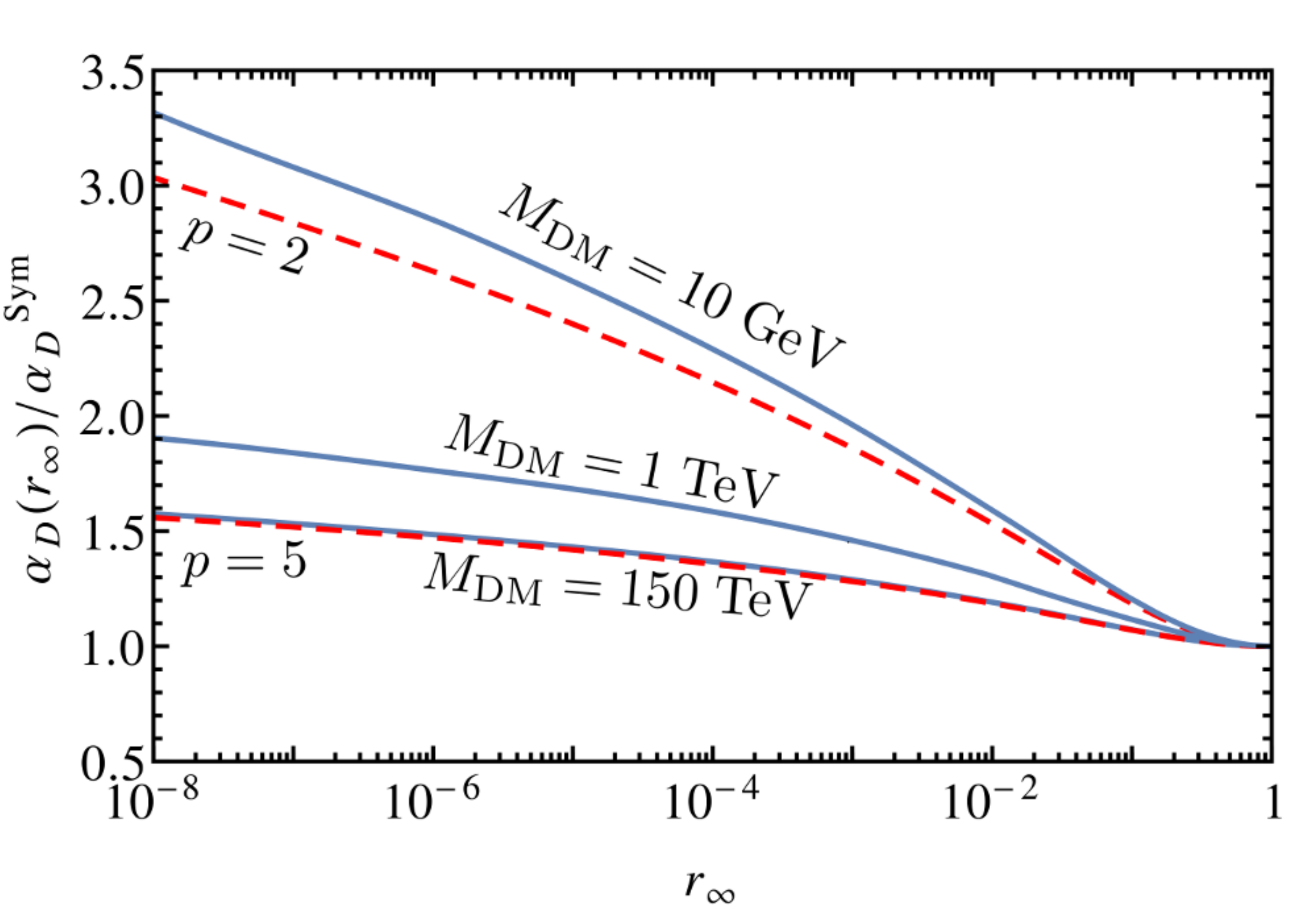}
\caption[]{\label{fig:ScalarMed_AlphaRatios} 
The equivalent of fig.~\ref{fig:VectorMed_AlphaRatios}, for DM coupled to a light scalar mediator. The annihilation cross-section around the time of freeze out scales as $\sigma_{\rm ann}^{\rm FO} \propto \alpha_D^p$, with $p=2$ and $p=5$ for small and large $M_{\rm DM}$ respectively.}
\end{figure}

The evolution of the fractional asymmetry $r$ is determined by \cref{eq:dr/dz}, if we identify
\begin{subequations}
\begin{align}
\sigma_* &\to \s_1 \,,
\label{eq:ScalarMed_sigma*}
\\
F(\xD)   &\to  \< \vrel^2 \ S_\ann^{(1)}(\zeta) \> 
= \aD^2 \, \< S_\ann^{(1)}(\zeta) /\zeta^2 \>  \,,
\label{eq:ScalarMed_F}
\end{align}
where the thermal average of \cref{eq:ScalarMed_F} is
\beq
F(\xD) = \frac{8}{\sqrt{\pi}\xD} 
\int_0^\infty du \ S_\ann^{(1)} (\sqrt{\zD/u}) \ u^{3/2} \, \exp(-u) \,.
\eeq 
\end{subequations}
We compute the coupling $\aD$ required to obtain the observed DM relic density, as a function of the asymmetry, and present the results in \cref{fig:ScalarMed_coupling}.

As in \cref{sec:VectorMed} for a vector mediator, we discern the perturbative and the Sommerfeld-enhanced regimes. In the perturbative regime, $F(\xD^\FO) \simeq \<\vrel^2\>_\FO = 6 / \xD^\FO$, $\Phi \simeq \Phi_\sym \simeq 3\sqrt{g_*^\FO}/(\xD^\FO)^2$, and $\sigma_* \Phi \propto \aD^2$. 
On the other hand, well within the Sommerfeld-enhanced regime, 
$F(\xD^\FO) \simeq 2\aD^3 \sqrt{\pi \xD^\FO}$,  
$\Phi \simeq \aD^3 \sqrt{\pi g_*^\FO / \xD^\FO}$ and $\sigma_* \Phi \propto \aD^5$. 
Then, from \cref{eq:AlphaRatio}, we estimate the ratio $\aD/\aD^\sym$ required to establish a fractional asymmetry $\rinf$, by setting $p=2$ and $p=5$ for small and large $\MDM$ respectively. We present the numerical solution for $\aD/\aD^\sym$ in \cref{fig:ScalarMed_AlphaRatios}, and compare it with the analytical approximation.

\section{Annihilation signals \label{Sec:AnnihilationSignals}}

The residual symmetric DM component may give rise to DM annihilation that could have observable implications~\cite{Graesser:2011wi}. Provided that the DM annihilation products include or cascade down to SM states, 
the DM annihilation inside haloes today may yield detectable signals~\cite{Bell:2014xta,Murase:2016nwx}. 
while the DM annihilation at late cosmic times may be constrained by CMB observations~\cite{Lin:2011gj}. 
In this section, we parametrise the expected signal strength, and estimate it in the Coulomb limit for DM annihilation in the Milky Way. The annihilation signals for a wide range of mediator masses, as well as observational constraints from different probes, will be presented elsewhere.

\medskip

\afterpage{

\begin{figure}[t]
\centering{\bf Vector mediator} 
\includegraphics[height=4.9cm]{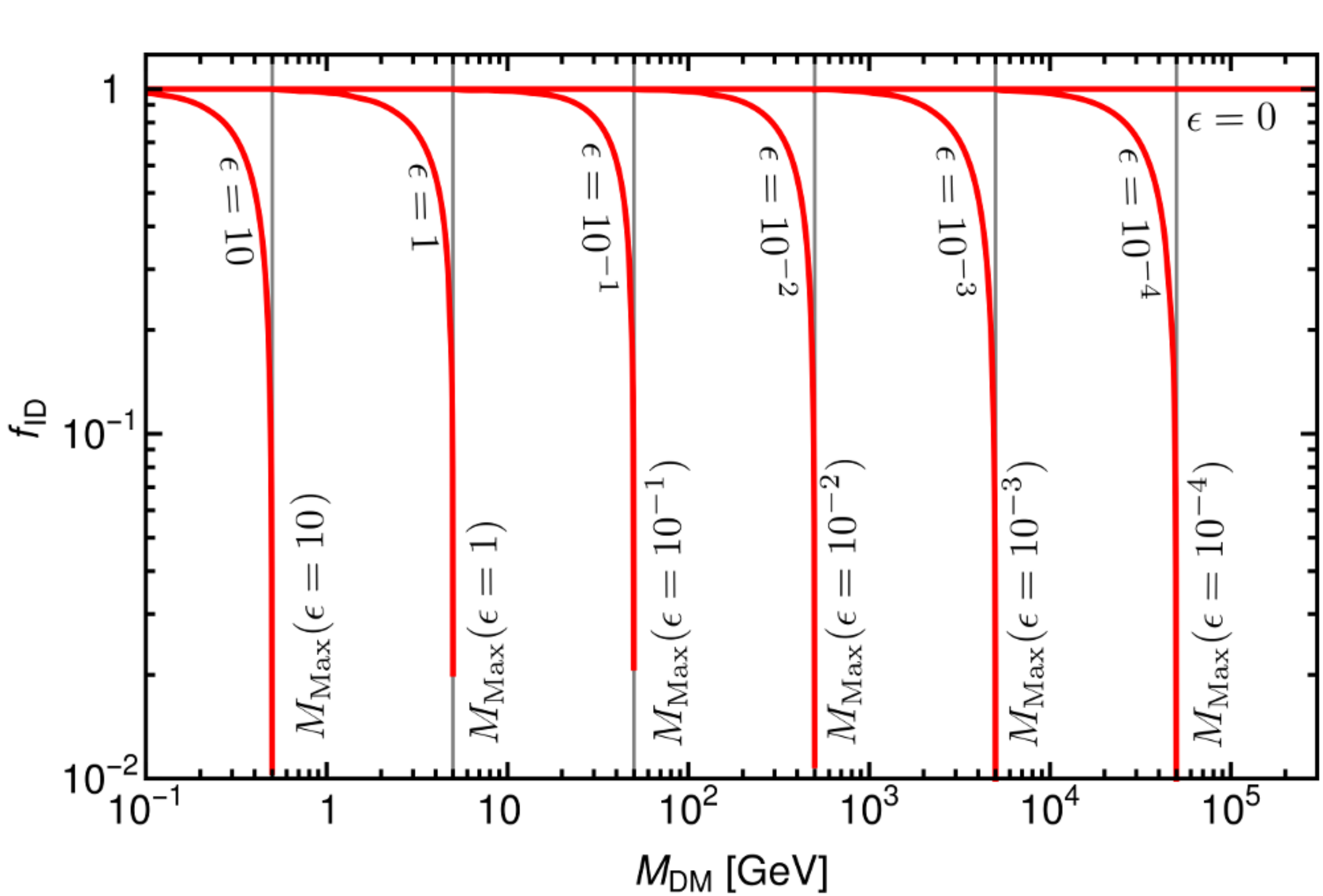}~~
\includegraphics[height=4.9cm]{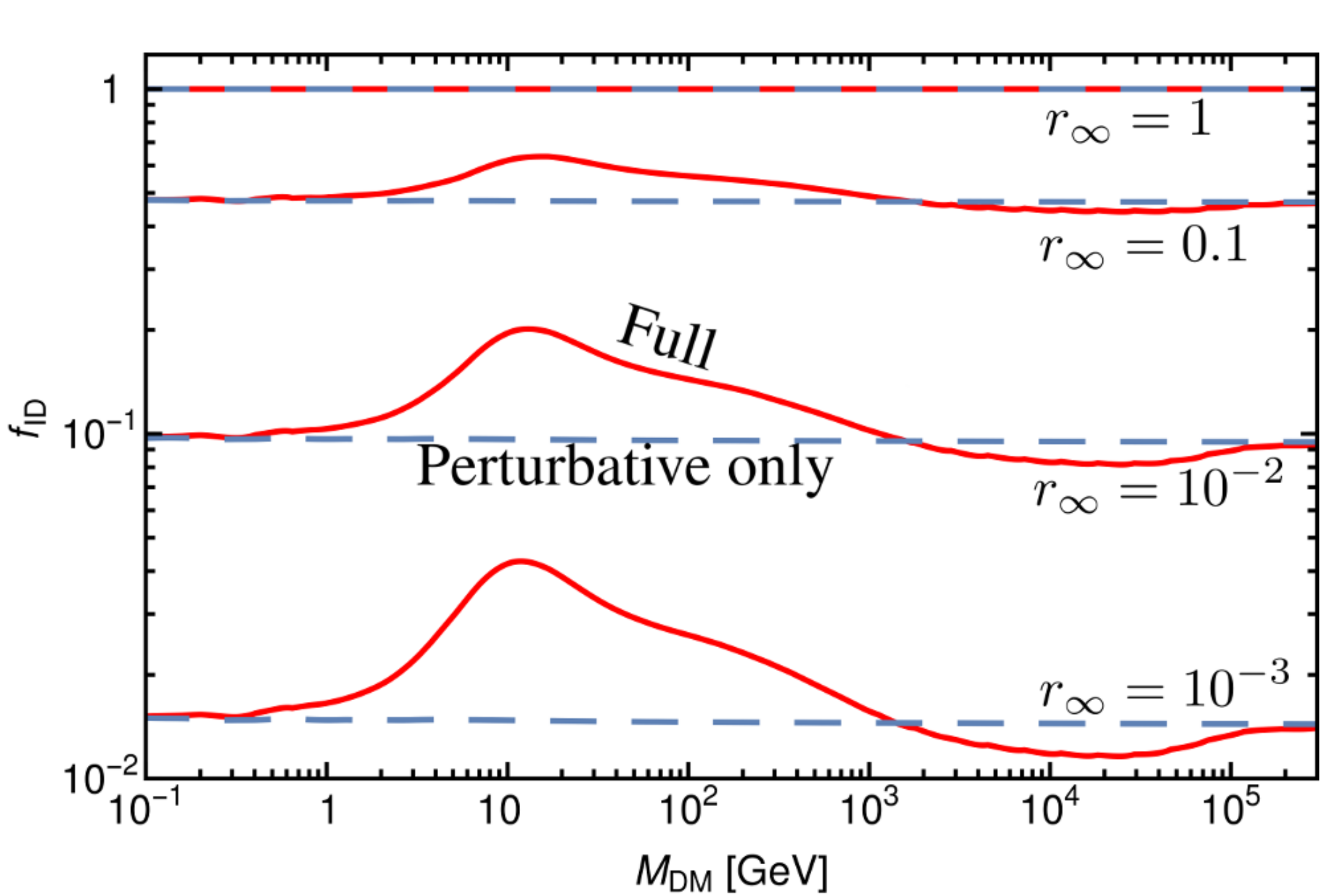}

\medskip

\includegraphics[height=4.9cm]{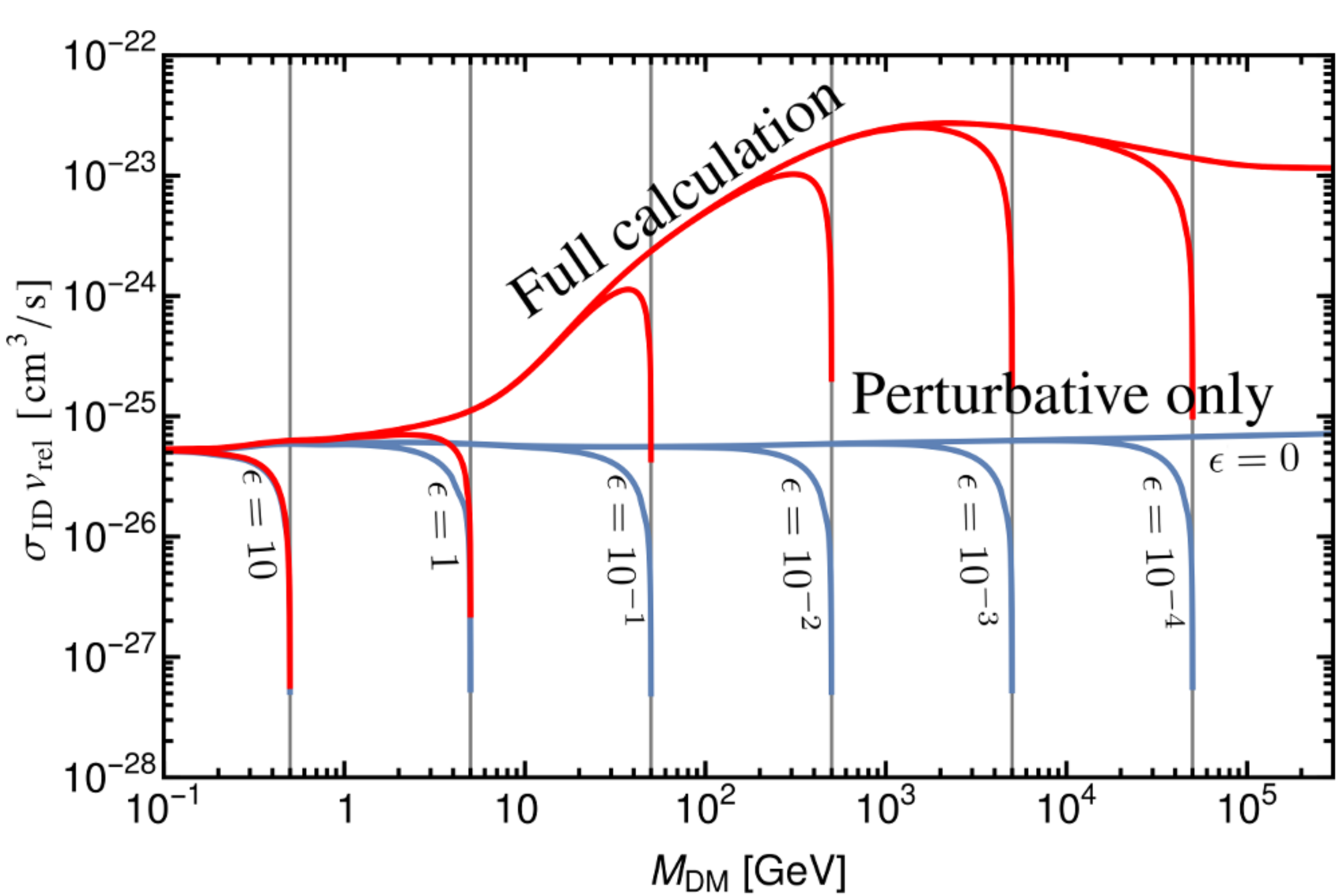}~~
\includegraphics[height=4.9cm]{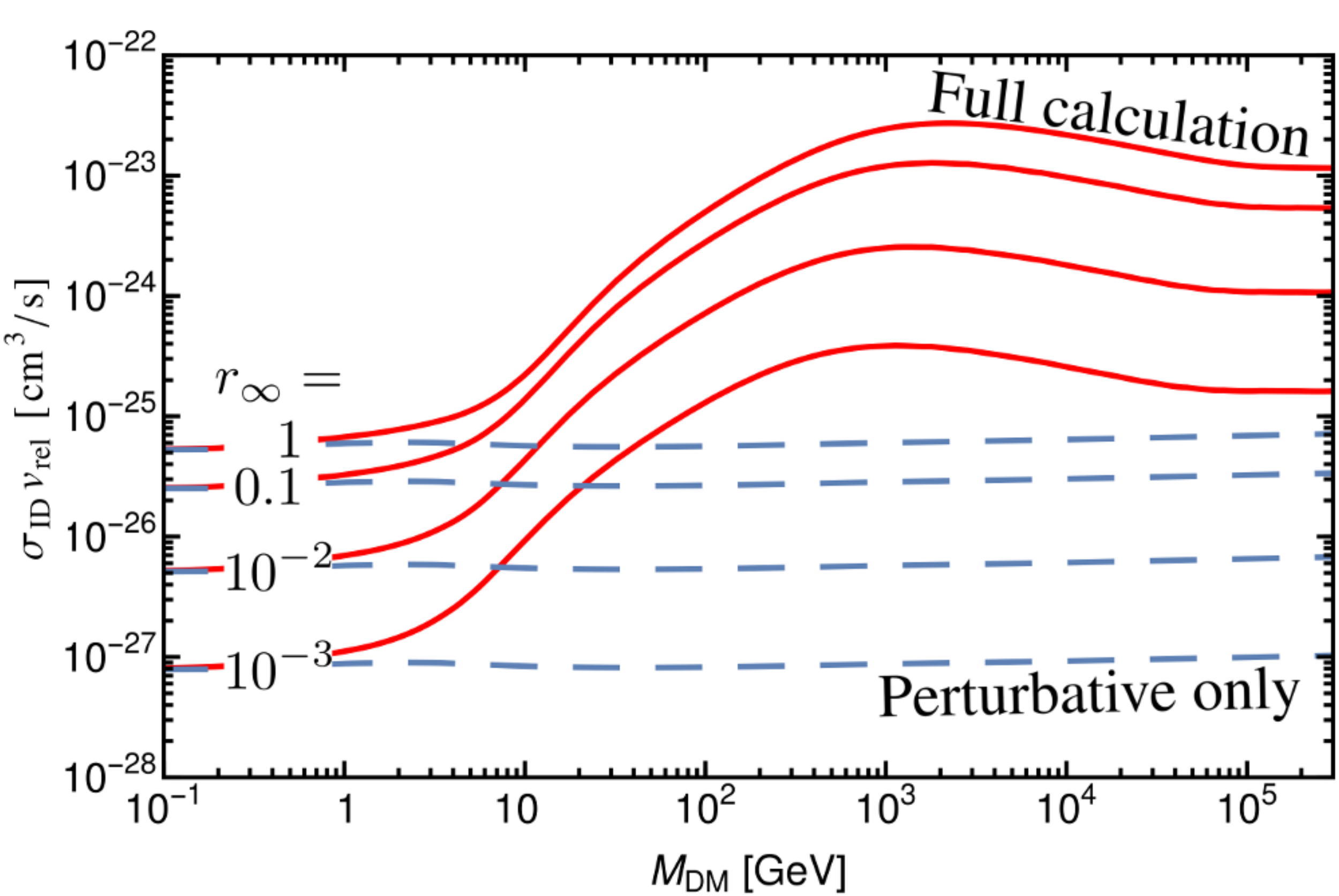}
\caption[]{\label{fig:VectorMed_AnnSignals}
\emph{Top:} The suppression factor of the expected annihilation signals with respect to symmetric DM of the same mass, $f_{\rm ID}$ vs $M_{\rm DM}$, for fixed $\epsilon = \eta_D^{}/\eta_B^{}$ (\emph{left}), and for fixed $r_{\infty}$ (\emph{right}).
\emph{Bottom:} The effective cross-section for indirect detection signals, 
$\sigma_{\rm ID}^{} v_{\rm rel} = [4\rinf/(1+\rinf)^2]\,\sigma_\inel v_{\rm rel}$ vs $M_{\rm DM}$, for fixed $\epsilon = \eta_D^{}/\eta_B^{}$ (\emph{left}), and for fixed $r_\infty$ (\emph{right}). 
In all panels, we have used $v_{\rm rel} = 10^{-3}$, which is relevant for indirect searches in the Milky Way, and evaluated the cross-sections in the Coulomb limit, which is a satisfactory approximation within a large range of values of the mediator mass.}
\end{figure}

\begin{figure}[t]
\centering{\bf Scalar mediator} 
\includegraphics[height=4.9cm]{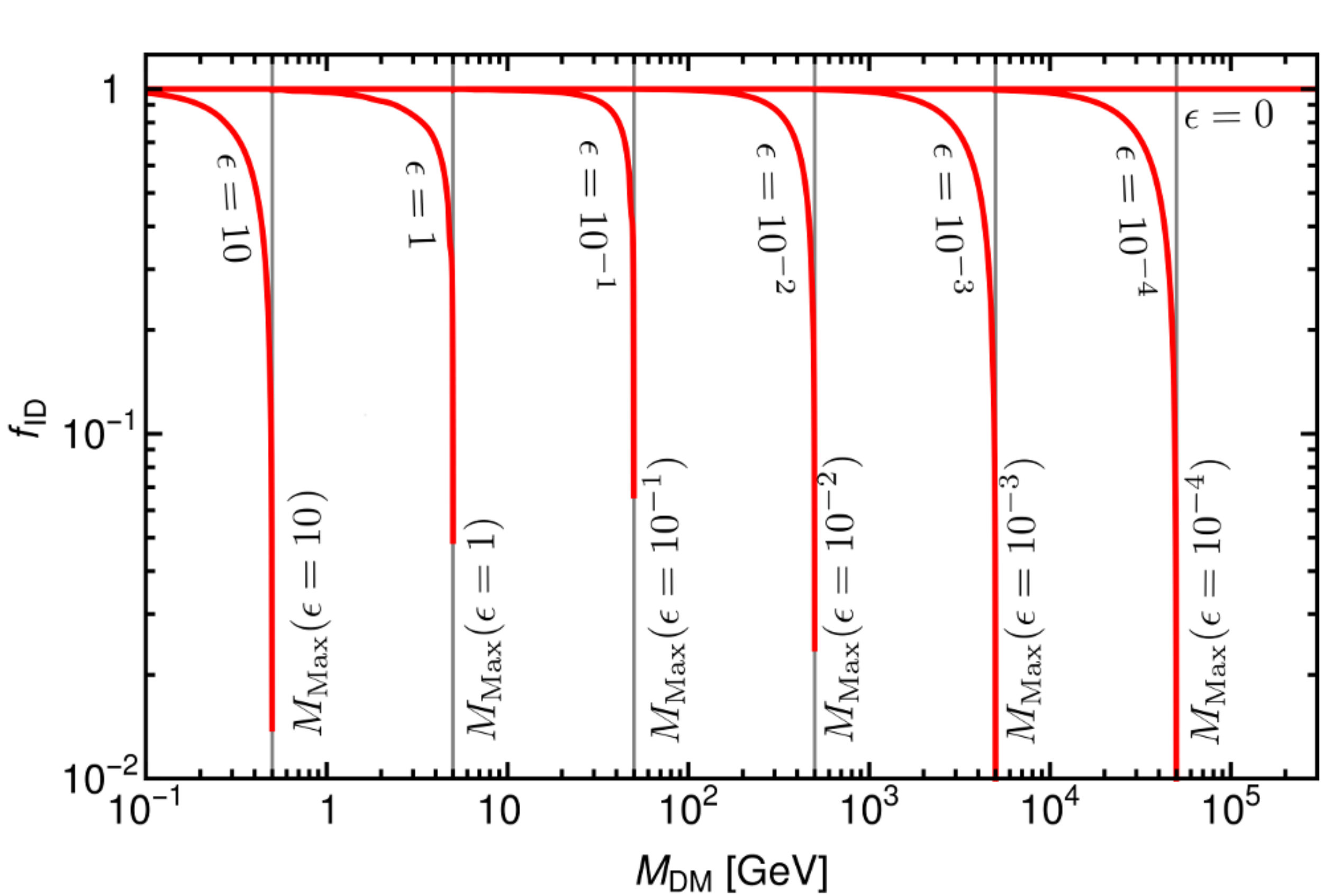}~~
\includegraphics[height=4.9cm]{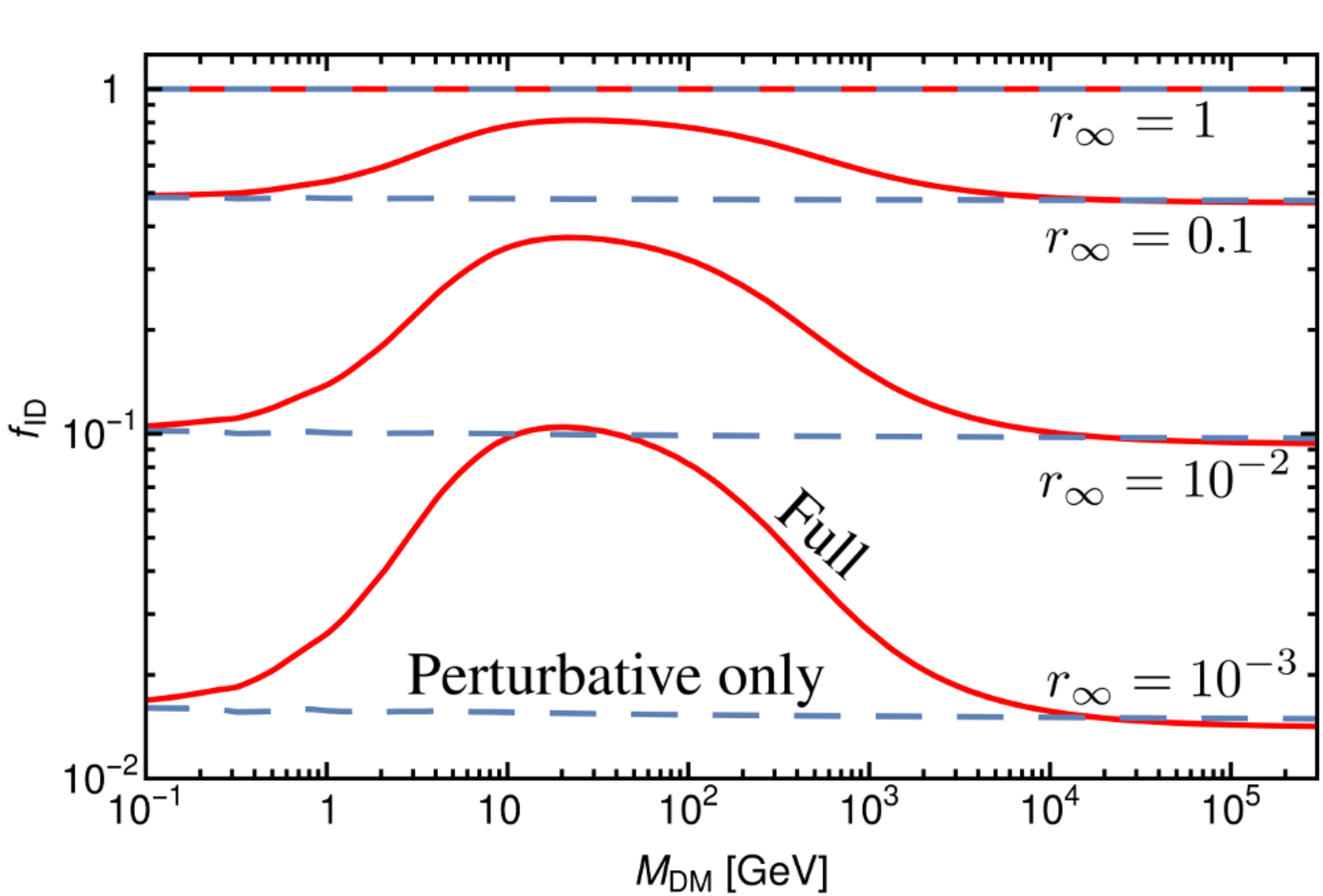}

\medskip

\includegraphics[height=4.9cm]{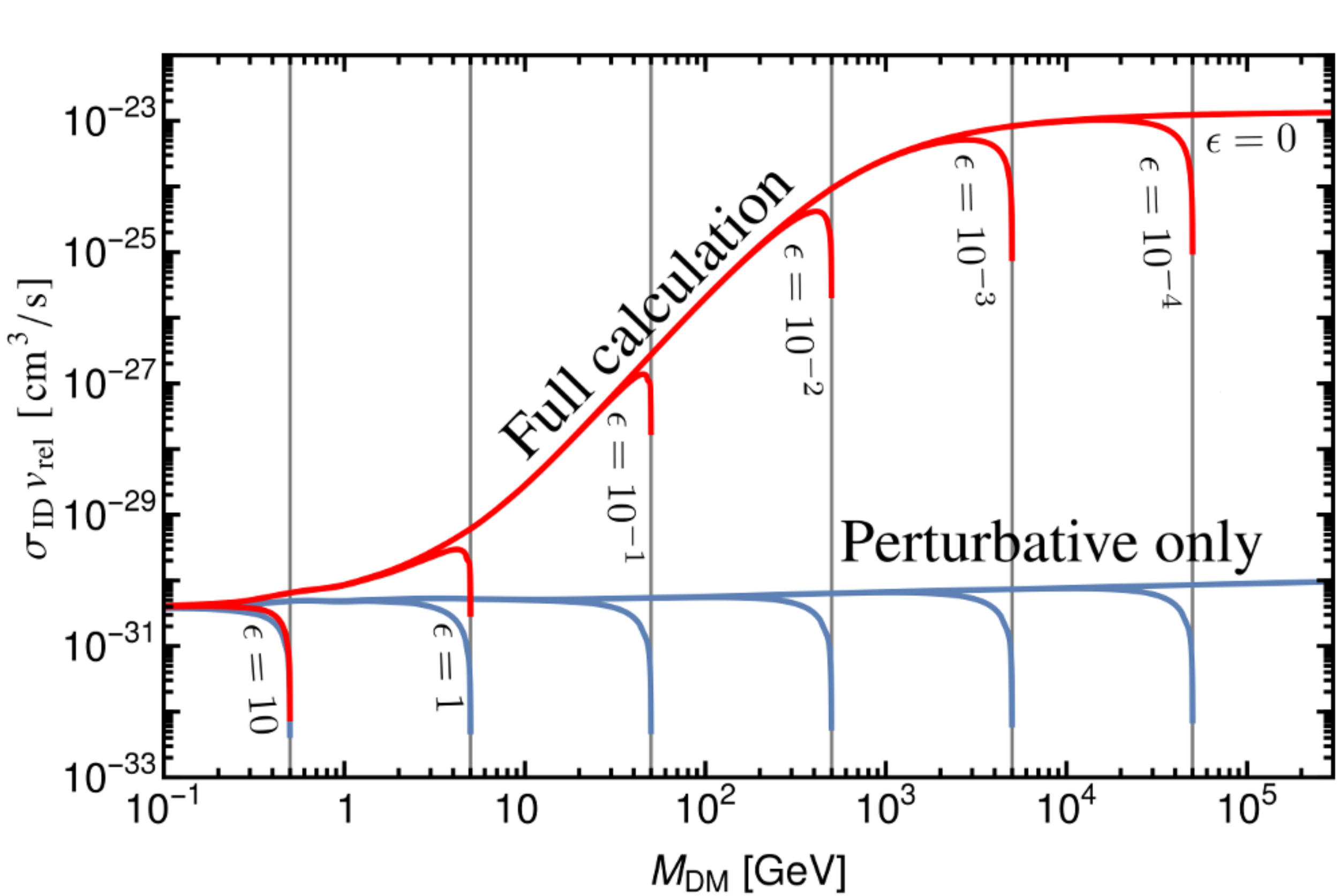}~~
\includegraphics[height=4.9cm]{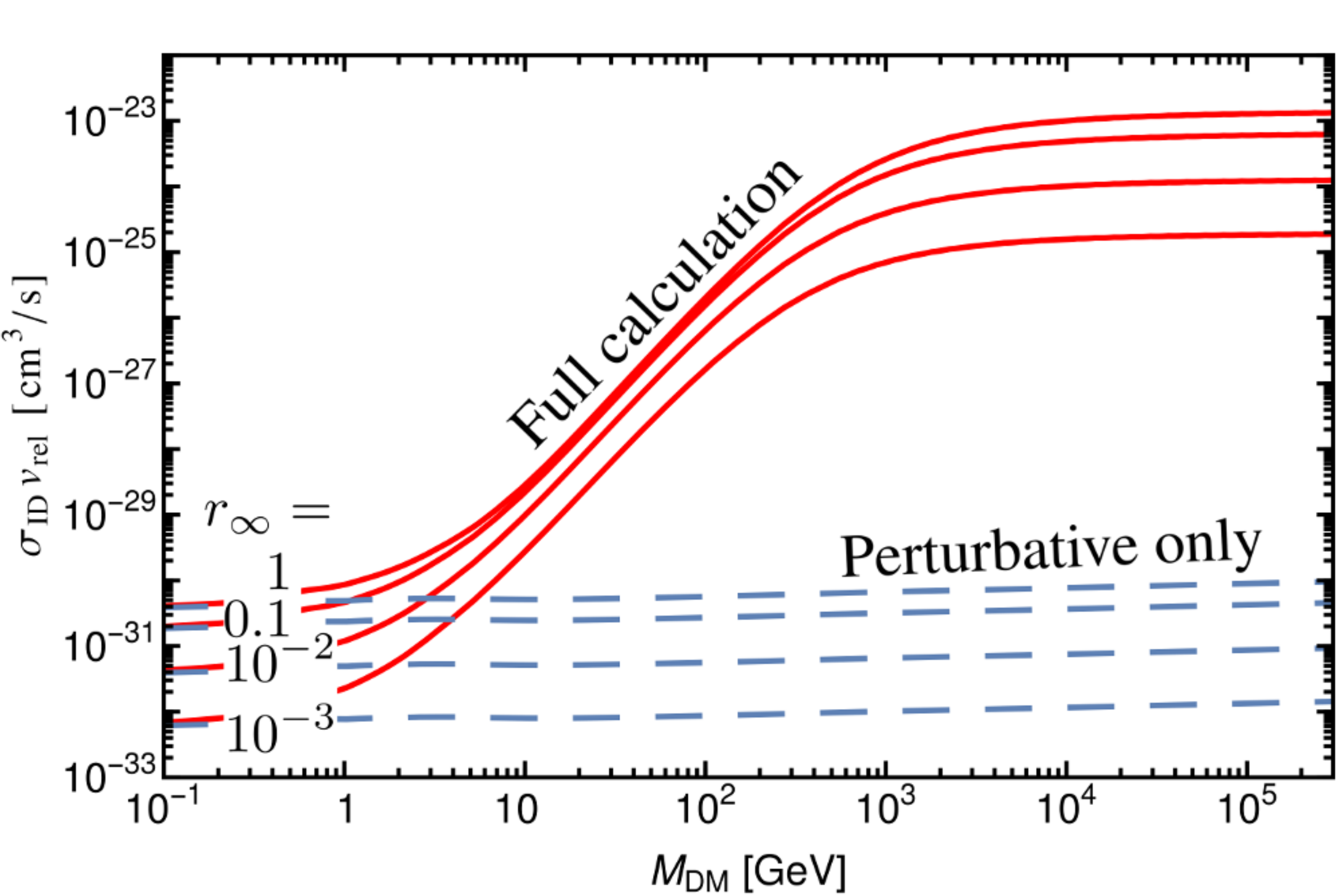}

\caption[]{\label{fig:ScalarMed_AnnSignals}
The equivalent of fig.~\ref{fig:VectorMed_AnnSignals}, for DM coupled to a light scalar mediator.}
\end{figure}
}

\subsection{Signal strength \label{sec:AnnSignalStrength}}

The asymmetric DM annihilation rate is suppressed with respect to symmetric DM due to the depleted population of dark antiparticles, albeit this suppression is ameliorated by the larger annihilation cross-section. For asymmetric DM, the expected signal rate is proportional to 
$Y_\infty^+ \, Y_\infty^- \<\sigma_\inel \vrel\> =  
[\etaD^2 r/(1-r)^2] \<\sigma_\inel \vrel\>$; for symmetric DM, it is proportional to $(Y_\infty^\sym)^2 \<\sigma_\inel^\sym \vrel\>$. 
Using \cref{eq:MDM} to express $\etaD = \epsilon \etaB$ in terms of $\rinf$, and noting that for symmetric DM, 
$\OmegaDM = 2Y_\infty^\sym \MDM \OmegaB/(\etaB m_p)$, we find that the suppression factor of the annihilation signals arising from asymmetric DM with respect to symmetric DM of the same mass is~\cite{Graesser:2011wi}
\begin{subequations}
\label{eq:fID}
\label[pluralequation]{eqs:fID}
\beq
f_{\mathsmaller{\rm ID}} = \dfrac{4\rinf}{(1+\rinf)^2} \,
\(\dfrac{\sigma_\inel \vrel}{\sigma_\inel^\sym \vrel}\) \,.
\tag{\ref{eq:fID}}
\label{eq:fID_def}
\eeq
Here, $\sigma_\inel$ includes all inelastic processes that contribute to the DM annihilation. 
Using \cref{eq:rinf_final_approx}, and assuming that the ratio $\s_\inel/\s_\inel^\sym$ is the same during freeze-out and at the velocities relevant for indirect detection,\footnote{%
This is a good approximation provided that both the freeze-out and the emission of the annihilation signals happen either well within the perturbative regime or well within the Sommerfeld-enhanced regime, for the couplings that correspond both to the symmetric and the asymmetric cases. } 
we obtain the analytical estimate for $f_{\mathsmaller{\rm ID}}$ in terms of $\rinf$ or $\epsilon$ and $\MDM$,
\beq
f_{\mathsmaller{\rm ID}} 
\approx \frac{2\rinf \ln(1/\rinf)}{1-\rinf^2} 
= \frac{1}{2} \[\frac{1 - \epsilon^2 \,\MDM^2 /(5~\GeV)^2}{\epsilon \,\MDM /(5~\GeV)} \]
\ln\[\dfrac{1+\epsilon \, \MDM/(5~\GeV)}{1-\epsilon \, \MDM/(5~\GeV)}\] ,
\label{eq:fID_approx}
\eeq 
\end{subequations}
where in the second step we used \cref{eq:MDM}.

For convenience, we define an effective cross-section for estimating the indirect detection signals of asymmetric DM, that can be directly compared to the annihilation cross-section of symmetric DM of the same mass,
\beq 
\sigma_{\mathsmaller{\rm ID}} \, \vrel \equiv 
\frac{4\rinf }{ (1+\rinf)^2} \ \sigma_\inel \, \vrel \,.
\label{eq:sigmaID}
\eeq

\subsubsection*{Coulomb regime}

The Coulomb approximation is suitable where the condition \eqref{eq:CoulombRegime} is satisfied; this encompasses a large range of mediator masses that yield observable signals from the Milky Way and the Dwarfs~\cite[fig.~2]{Cirelli:2016rnw}. We present the numerical evaluation of $f_{\mathsmaller{\rm ID}}$ and $\sigma_{\mathsmaller{\rm ID}} \vrel$  in \cref{fig:VectorMed_AnnSignals,fig:ScalarMed_AnnSignals}, using the numerical computation of $\rinf$ presented in \cref{Sec:AsymFO}, and the formulae provided in \cref{Sec:SommerfeldAsymFO} for the inelastic cross-sections in the Coulomb regime. We adopt the indicative value $\vrel = 10^{-3}$ for the relative velocity of the DM particles, which is a typical value for the Milky Way. We discern the following regimes: 
\bit
\item
For  $\MDM \lesssim 10$~GeV, the DM annihilation occurs mostly in the perturbative regime, both during the DM chemical decoupling in the early universe, and inside haloes today; $\sigma_{\mathsmaller{\rm ID}} \, \vrel$ is fairly independent of $\MDM$. 
\item In the intermediate mass range, $\MDM \sim 10~\GeV - 1~\TeV$, the Sommerfeld effect has negligible impact during freeze-out, but is significant inside galaxies today, where the average velocity is lower. As above, this sets $\sigma_0$ or $\sigma_1$ to be nearly independent of $\MDM$, or equivalently $\aD \propto \MDM$. Since 
$S_\ann^{(0)}, \ S_\BSF \propto \aD$ and $S_\ann^{(1)} \propto \aD^3$ at $\aD \gtrsim \vrel$ [cf.~\cref{eq:VectorMed_Sfactors,eq:S_ann_1}], the effective cross-section for indirect detection signals scales as 
$\sigma_{\mathsmaller{\rm ID}} \, \vrel \propto \MDM$~and~$\MDM^3$, for fixed $\rinf$, for a vector and a scalar mediator respectively. 
\item For $\MDM \gtrsim 1~\TeV$, the Sommerfeld effect is operative both during freeze-out and inside haloes today, albeit the enhancement is different due to the different velocity. Consequently, $\sigma_{\mathsmaller{\rm ID}} \, \vrel$ becomes again insensitive to $\MDM$, but is significantly larger than what expected from perturbative annihilation, by a factor of $\vrel^\FO/\vrel$ and $(\vrel^\FO/\vrel)^3$ for a vector and scalar mediator respectively.
\eit

It is notable that, due to the Sommerfeld enhancement, $\sigma_{\mathsmaller{\rm ID}} \, \vrel$ can be larger, even by many orders of magnitude, than the cross-section for symmetric DM annihilating via contact interactions, for a large range of masses, $\MDM \gtrsim 10~\GeV$, and highly asymmetric DM. For the Milky Way, this includes $\rinf$ as low as $10^{-3}$ for a vector mediator and $10^{-8}$ for a scalar mediator. For probes where the DM velocity dispersion is lower, such as the Dwarf galaxies and the CMB, the range of $\rinf$ that yields significant annihilation signals could extend to even lower values. 

However, as is well known, at very low velocities, the Coulomb approximation fails. Then, the mediator mass affects the Sommerfeld effect and therefore the expected annihilation rate, as we describe below.

\subsubsection*{Outside the Coulomb regime}

At low velocities, $\vrel (\MDM/2) \lesssim m_{\rm med}$, the Coulomb approximation fails. This regime is relevant for DM annihilation during CMB, for any mediator mass that allows decay into SM charged fermions ($m_{\rm med} \gtrsim \MeV$) and DM mass below the unitarity limit ($\MDM \lesssim 300~\TeV$, cf.~\cref{Sec:Unitarity}). Within more limited parameter space, the DM annihilation inside haloes today is also outside the Coulomb regime~\cite[fig.~2]{Cirelli:2016rnw}. 

Outside the Coulomb regime, the features of the inelastic cross-sections change in two important ways: (i) The cross-sections exhibit parametric resonances, at discrete values of $\aD \MDM / m_{\rm med}$, that correspond to the thresholds for the existence of bound states. (ii) The velocity scaling of the inelastic cross-sections, both on- and off-resonance, depends on the partial waves $\ell$ of the initial-state wavefunction that participate in the process;
at $\vrel (\MDM/2) \ll m_{\rm med}$, the scaling becomes $\sigma_\inel \vrel \propto \vrel^{2\ell}$.
This implies, among else, that the dominant inelastic processes at low velocities may be different than those in the Coulomb regime. Detailed studies of the inelastic cross-sections can be found in Refs.~\cite{Petraki:2016cnz, An:2016gad, An:2016kie}.

\subsection{Indirect probes}

For a vector mediator, the CMB observations are expected to be broadly more constraining than the $\gamma$-ray observations of DM haloes, as was recently shown in Refs.~\cite{Cirelli:2016rnw, Bringmann:2016din} for symmetric DM. 
The strength of the CMB constraints is in part due to the fact that the DM annihilation into two vector bosons is an $s$-wave process. For $s$-wave processes, the Sommerfeld enhancement of the annihilation signals  is greater during CMB than inside haloes today, since the DM velocity is much lower during CMB. On the other hand, BSF with emission of a vector boson --   a $p$-wave process -- is negligible during CMB, but strengthens significantly the $\gamma$-ray constraints from the Milky Way and the Dwarf galaxies~\cite{Cirelli:2016rnw}. 
Moreover, antiproton observations of the Milky Way probe parameter space that is not constrained by CMB~\cite{Cirelli:2016rnw}.

For a scalar mediator, the direct annihilation is to leading order a $p$-wave process, and is therefore  insignificant during CMB. However, it has been recently pointed out that BSF with emission of a scalar mediator -- which is subdominant with respect to annihilation in the Coulomb regime -- contains an $s$-wave component, albeit suppressed by $\aD^2$ in comparison to annihilation into vector mediators. 
This yields CMB constraints, even though less stringent than those for a vector mediator~\cite{An:2016kie}.
It is plausible that for a scalar mediator, the $\gamma$-ray observations of the Milky Way and the Dwarf galaxies are more constraining than CMB (both for symmetric and asymmetric DM), since for the corresponding velocities, the DM annihilation is described by the Coulomb limit along a significant range of mediator masses. Note though that even in the Coulomb limit, the $p$-wave annihilation is suppressed, albeit by an extra factor of $\aD^2$ rather than $\vrel^2$, as noted in \cref{sec:ScalarMed}. 

\smallskip

CMB constraints on asymmetric DM have been previously deduced in Ref.~\cite{Lin:2011gj}, which considered DM annihilating via $s$-wave contact interactions. Indicatively we mention that, using WMAP7 data, Ref.~\cite{Lin:2011gj} excluded 
$\rinf > 3 \times 10^{-4}$ for $\MDM \approx 100~\MeV$, and 
$\rinf > 10^{-1}$ for $\MDM \approx 8~\GeV$, assuming the efficiency of the annihilation products in  ionising the medium was maximal. The results of the present work show that for asymmetric DM coupled to light mediators, the constraints should be anticipated to strengthen very significantly. In particular, observations could constrain a larger $\MDM$ range and lower $\rinf$ values, as well as probe potentially both vector and scalar mediators.

\subsection{Decay of the force mediators}

The detectability of the radiative signals produced in the annihilation of DM into dark mediators relies on a coupling of the dark sector with the SM. A dark photon may mix kinetically with hypercharge, $\d {\cal L}_{V} = -(\kappa/2) \: \FD \FY$, while a scalar mediator may mass mix with the SM Higgs. For $\mVD, \, m_\vf > 1.022~\MeV$, these mixings lead to the decay of the dark mediators into SM charged fermions. 

The corresponding decay rates at rest are (see e.g.~\cite{Batell:2009yf,Kainulainen:2015sva})
\begin{subequations}
\begin{align}
\Gamma_{\VD} 
&\simeq \sum_f \frac{q_f^2}{3} \, \kappa^2 \aEM \ \mVD \ \(1- 4m_f^2 / \mVD^2\)^{3/2} \,, 
\label{eq:DecayRate_VD} 
\\
\Gamma_{\vf}
&\simeq \sum_f \frac{m_f^2}{8\pi \vEW^2 } \sin^2\b \ m_\vf \ \(1- 4m_f^2 / \mVD^2\)^{3/2} \,, \label{eq:DecayRate_phi}
\end{align}
\end{subequations}
where $q_f$, $m_f$ are the electric charges and masses of the SM fermions, $\sin\b$ is the hypothesised mixing between the dark scalar and the Higgs, and  $\aEM =1/137$, $\vEW = 246~\GeV$. 
The requirement that the cosmological abundance of the dark mediators, if significant, decays sufficiently before BBN sets a lower bound on the couplings to the SM. While the precise bound depends on the mass of the mediator and the spectrum of the decay products, it could be as stringent as $\G \gtrsim (0.03\,\sec)^{-1}$ (see e.g.~\cite{Berger:2016vxi}).

However, the mediators produced in the DM annihilation and BSF processes are boosted by a factor $\sim (\MDM/m_{\rm med})$; thus, the above requirement does not automatically ensure that they decay in time scales that are relevant to our indirect probes. Including this boost factor, the dark mediator decay lengths are
\begin{subequations}
\begin{align}
\tau_{\mathsmaller{\VD}} \times (\MDM / \mVD) &\simeq 
0.26~\pc \times \(\frac{1}{\sum_f q_f^2}\) \ \(\frac{10^{-10}}{\kappa}\)^2 \, \(\frac{\MDM}{\TeV}\) \, \(\frac{\MeV}{\mVD}\)^2 \,, 
\label{eq:tau_VD}
\\
\tau_{\vf} \times (\MDM / m_\vf) &\simeq 
37~\pc \times \[\frac{(511~\keV)^2}{\sum_f m_f^2}\] \ \(\frac{10^{-6}}{\sin \b}\)^2 \, \(\frac{\MDM}{\TeV}\) \, \(\frac{\MeV}{\mVD}\)^2 \,.
\label{eq:tau_phi}
\end{align}
\end{subequations}
This implies that the decay of the mediators is prompt in galactic scales, as well as during CMB, even for very light dark mediators that give rise to long-range scattering [cf.~condition~\eqref{eq:long-range}], and couplings to the SM that are (well) below the current constraints from direct detection experiments and other probes. For more details, see e.g.~\cite{Cirelli:2016rnw,Bringmann:2016din,Kainulainen:2015sva}.

\subsection{Cosmological caveat}

\Cref{eq:fID,eq:sigmaID} do not always suffice to predict the indirect detection signals that should be expected from asymmetric DM annihilation. Cosmological events following the DM chemical freeze-out may further suppress the fraction of DM that is available to participate in the annihilation processes at later times. This is, in fact, particularly relevant to scenarios of asymmetric DM coupled to light or massless force mediators.

If DM bares a particle-antiparticle asymmetry and couples to a light or massless dark photon, then gauge invariance mandates that a second dark species, charged under the same dark force, bares an asymmetry and has survived until today, such that the total dark electric charge of the universe vanishes.\footnote{%
This is, of course, analogous to ordinary protons and electrons in the SM, and to mirror electrons and mirror protons in mirror matter models~\cite{Foot:2014mia}. For other models of atomic DM, see e.g.~\cite{Kaplan:2009de,Petraki:2011mv,vonHarling:2012yn,Foot:2014uba,Choquette:2015mca,Agrawal:2017rvu}.}   
Indeed, assuming that inflation diluted any pre-existing particle content of the universe, 
and that all processes that have taken place since then, including the decay of the inflaton, are governed by a gauge-invariant theory, the gauge-charge neutrality of the universe follows. 
While this evidently implies a vanishing dark electric charge in the case of an unbroken gauged $U(1)_D$ and a massless dark photon, a similar conclusion holds even for a mildly broken $U(1)_D$ and a sufficiently light dark photon~\cite{Petraki:2014uza}.

The presence of a second asymmetric dark species with the opposite net $U(1)_D$ charge implies that \emph{stable} atomic bound states may form, thus trapping dark particles that would be otherwise available to participate in annihilation processes with the residual dark antiparticles. 
The efficiency of dark recombination -- the formation of dark atoms in the early universe -- depends on the masses of both species, which implies that \cref{eq:sigmaID} is insufficient to predict the expected DM annihilation signals.\footnote{%
As discussed in \cref{Sec:DStemperature}, the additional degrees of freedom in the dark sector will affect $\TD$ and hence also the $\langle \sigma_{\rm ann}v_{\rm rel} \rangle$ required to obtain the observed DM abundance. To a good approximation, the required annihilation cross-section scales as $\<\s_\ann\vrel\> \propto \TD^\FO/\TSM^\FO$.
}

Asymmetric DM that couples to a light scalar mediator can form stable bound states due to the attractive nature of the interaction between particles of the same species (as well as between particles and antiparticles). In contrast to the vector mediator case, the formation of particle-particle bound states does not neutralise the interaction mediated by the scalar. This may result in the cosmological formation of a spectrum of multiparticle bound states~\cite{Wise:2014jva}, 
which of course modifies the number density of the DM states, as well as their annihilation cross-section. In this event, the estimation of the DM annihilation signals based solely on \cref{eq:sigmaID} would be inaccurate.

The comprehensive computation of the annihilation signals expected in these scenarios, and the resulting observational constraints, merit dedicated analyses that will employ the results presented here, but are beyond the scope of the present work. It is worth noting that in these scenarios, signals for indirect DM searches may also arise from the radiative formation, or the excitation and de-excitation of the stable DM bound states --- in particular, from the formation of dark atoms~\cite{Pearce:2015zca} or other related transitions~\cite{Cline:2014eaa} in the case of a vector mediator, and from the formation of dark particle-particle bound states in the case of a scalar mediator~\cite{Pearce:2013ola}. 

\smallskip

We note in passing that the above cosmological considerations are important also in determining other phenomenological aspects of the scenarios considered here, and in particular the DM self-interactions inside haloes today. The formation of stable bound states typically screens or curtails the DM self-scattering. It is then essential that phenomenological studies take into account the entire cosmological history, of which the DM chemical freeze-out computed here is essentially the first part.


\section{Unitarity limit \label{Sec:Unitarity}}

\subsection{Long-range vs. contact-type interactions \label{sec:Unitarity_general}}

Partial-wave unitarity sets an upper limit on the 2-to-2 total inelastic cross-section. In the non-relativistic regime, the unitarity limit reads~\cite{Griest:1989wd}
\beq
\sigma_\inel^{(J)} \, \vrel \leqslant \sigma_{\rm uni}^{(J)} \, \vrel = \frac{4\pi (2J+1)}{\MDM^2\,\vrel} \,,
\label{eq:sigmaUni}
\eeq
where $J$ denotes the partial wave. Physically, the above limit corresponds to the saturation of the probability for 2-to-2 inelastic scattering to 1. Within a given theory, it is typically expected to be realised or approached either at sufficiently large couplings or near resonant points, if at all.

Assuming that DM has thermalised in the early universe, the limit \eqref{eq:sigmaUni} implies that DM cannot be too heavy, because it would not have annihilated sufficiently, down to its observed density. This consideration was first put forward in Ref.~\cite{Griest:1989wd}, to obtain an upper bound on the mass of symmetric thermal-relic DM, which amounts to $\sim 83~\TeV$ for non-self-conjugate DM, using the current measurement of $\OmegaDM$. However, two refined considerations relating to the velocity dependence of $\sigma_\uni \vrel$ and the contribution of high partial waves to the inelasticity, raise this upper bound on $\MDM$ considerably. We now expound on these points.

\subsubsection[The velocity scaling of $\sigma_{\rm uni} \, \vrel$]{The velocity scaling of $\boldsymbol{\sigma_{\rm uni} \, \vrel}$ \label{sec:Unitarity_general_velocity}}

For the purpose of computing the upper bound on the mass of thermal relic DM, Ref.~\cite{Griest:1989wd} evaluated $\sigma_\uni \vrel$ at a fixed value of $\vrel$, of the order of those that occur during freeze-out. However, since freeze-out is not instantaneous, the proper computation of the DM relic density necessitates accounting for the variation of the thermally averaged $\<\sigma_\ann \vrel\>$ with the temperature if $\sigma_\ann \vrel$ depends on $\vrel$. Taking into account that $\<\sigma_\uni \vrel\>$ increases as the universe expands, raises the upper bound on the mass of non-self-conjugate symmetric DM annihilating via $s$-wave processes, to $M_{\uni,\sym}^{(J=0)} \approx 140~\TeV$~\cite{vonHarling:2014kha,Berger:2008ti}, assuming that DM annihilates into a thermal bath of the same temperature as the SM. While Ref.~\cite{Griest:1989wd} discussed the effect of the temperature dependence of $\<\sigma_\uni \vrel\>$ on $M_{\uni,\sym}$, the physical significance of the $\sigma_\uni$ scaling with $\vrel$ was not recognised, and the value of $M_{\uni,\sym}$ resulting from this computation was regarded as overly conservative.

More recently, it has been pointed out that the velocity dependence of $\sigma_\uni$ has physical significance, and therefore needs to be taken into account~\cite{vonHarling:2014kha}. In particular, the scaling $\sigma_\uni \vrel \propto 1/\vrel$ suggests that the unitarity limit may be realised only if DM annihilates via a long-range interaction, as we shall now discuss in more detail.

Let us begin by assuming that the unitarity limit can be realised by contact-type interactions. 
For an interaction mediated by a heavy force carrier of mass $m_{\rm med} \gtrsim \MDM$, the inelastic cross-section scales as $\sigma_\inel \vrel \sim \aD^2 \MDM^2 / m_{\rm med}^4$. Realising the unitarity limit \eqref{eq:sigmaUni} via such an interaction would then require a large coupling $\aD^\uni \sim (m_{\rm med}/\MDM)^2 / \sqrt{\vrel} \gtrsim m_{\rm med}/\MDM \gtrsim 1$. While the largeness of the required coupling is not surprising, recasting the above requirement implies $m_{\rm med} \lesssim \aD^\uni \MDM$. 
This condition states that the range of the interaction between two DM particles, $m_{\rm med}^{-1}$, is comparable or larger than their Bohr radius, $(\aD^\uni \MDM/2)^{-1}$, and marks the regime where the interaction manifests as long-range, thereby contradicting the original premise of a contact-type interaction. 
Including terms of higher order in $\aD$ from the perturbative expansion used to compute $\sigma_{\inel}$ would not essentially change this conundrum.  (See also Refs.~\cite{Hisano:2002fk, Hisano:2003ec} for a related discussion.)  Similarly, for $m_{\rm med} < \MDM$, an inelastic cross-section computed perturbatively would scale as $\sigma_\inel \vrel \sim \aD^2 / \MDM^2$; for example, in a dark QED theory, $\sigma_\ann^{\rm pert} \vrel = \pi \aD^2/\MDM^2$ [cf.~\cref{eq:sigma0}]. Realising the unitarity limit then requires $\aD^\uni \sim \text{few}/\sqrt{\vrel} \gtrsim 1$, which again implies $\aD^\uni (\MDM/2) \gtrsim m_{\rm med}$, and the above discussion follows.

Including the non-perturbative effects that arise from the long-range nature of an interaction in the regime $\aD (\MDM/2)\gtrsim m_{\rm med}$, results in the inelastic cross-sections exhibiting the same parametric dependence on $\MDM$ and $\vrel$ as the unitarity limit of \cref{eq:sigmaUni}. This is manifested by \cref{eq:VectorMed_sigmas,eq:sigma0,eq:VectorMed_Sfactors,eq:sigma_ann_ScalarMed}, and holds true at least for sufficiently large couplings, $m_{\rm med} < (\MDM/2) \vrel < (\MDM/2) \aD$, that are relevant for realising the unitarity limit. 
Therefore, requiring that the unitarity limit is realised in a theory with long-range interactions, simply yields a numerical value for a dimensionless coupling; for the inelastic cross-sections of \cref{sec:VectorMed,sec:ScalarMed},  this is $\aD^\uni \approx 0.85$ and $\aD^\uni \approx 1.4$ respectively (see also discussion in \cref{sec:Unitarity_general_PartialWaves}). 
Of course, the estimation of $\aD^\uni$ may be improved by including terms of higher order in $\aD$ in the perturbative expansions that have been employed in the computation of $\sigma_\inel$.\footnote{
Even with non-perturbative effects included, computations of interaction cross-sections in weakly-coupled theories typically employ perturbative expansions at various points. For example, the following well-known perturbative approximations have been used in computing the inelastic cross-sections of \cref{Sec:SommerfeldAsymFO}: 
(i) Only the leading order contribution to the two-particle interaction at infinity, the one-boson exchange diagram, is considered, and 
(ii) the energy exchange is neglected with respect to the momentum exchange, since it scales as $q^0 \sim {\bf q}^2/\mu$ and $|{\bf q}|$ scales with $\vrel$ and $\aD$, where $\mu$ is the reduced mass of the interacting particles; these two approximations lead to the Coulomb potential.
(iii) Only the leading order contribution to the radiative vertices is taken into account.
}
Indeed, the fact that $\sigma_\inel$ appears to excced $\sigma_\uni$ at $\aD > \aD^\uni$ in the cross-sections of \cref{Sec:SommerfeldAsymFO} affirms that higher-order perturbative corrections will affect the estimation of $\aD^\uni$ considerably. 
However, the necessity to include higher-order terms in $\aD$ in order for $\sigma_\inel$ to converge at or below $\sigma_\uni$ for any $\aD$, does not preclude the realisation of the unitarity limit, $\sigma_\inel \simeq \sigma_\uni$, in this regime of the theory.\footnote{
In any case, the precise value of $\aD^\uni$ in a given theory, if it exists, does not affect the determination of the unitarity bound on the mass of thermal relic DM, which depends only on $\sigma_\uni$ of \cref{eq:sigmaUni}. 
}
%
This is very different from the apparent violation of unitarity at $\aD > \aD^\uni$ in computations that assume a contact interaction, where, as we saw, the unitarity limit yields a comparison between scales that indicates new physical effects are at play.

Finally, as noted earlier, the unitarity limit on the inelastic cross-section may be realised at resonant points. Indeed, in the regime 
$(\MDM/2)  \vrel < m_{\rm med} < (\MDM/2)  \aD$, 
the inelastic cross-sections exhibit parametric resonances on the thresholds for the existence of bound states. Leading order computations (that include the Sommerfeld effect) suggest that these resonances may grow to be as large as the unitarity limit.  
However, the velocity dependance of the inelastic cross-sections around these resonant points is different than that of \cref{eq:sigmaUni}. 
Even if $\s_\inel \vrel$ is maximal at a specific $\vrel$, this does not hold at all velocities. 
Equivalently, even if $\<\s_\inel \vrel\>$ is nearly maximal at some temperature, it drops below the unitarity limit
as the universe expands. Therefore, the unitarity limit on the mass of thermal-relic DM cannot be reached.

\smallskip
Having discussed why $\sigma_{\inel} \approx \sigma_\uni$ and $\MDM \approx M_\uni$ may be realised only by interactions that manifest as long-range, in \cref{sec:Unitarity_MassBounds}, we extend the computation of the upper bound on the mass of thermal-relic DM to asymmetric DM, using the relic density calculations of the previous section that focused on long-range interactions.

\subsubsection{Higher partial waves  \label{sec:Unitarity_general_PartialWaves}}

Higher partial waves may contribute significantly to the total inelastic cross-section, and consequently to the depletion of DM in the early universe.

Reference~\cite{Griest:1989wd} argued that the $s$-wave contribution dominates the inelastic cross-section, since higher partial waves are suppressed in the non-relativistic regime by $\vrel^{2J}$. This is indeed true for contact interactions, provided that no symmetry eliminates the $s$-wave contribution. However, the unitarity limit cannot be realised by contact interactions. As discussed above, it may be realised by long-range interactions, which exhibit the same velocity scaling, $\sigma_\inel^{(J)} \vrel \propto 1/\vrel$, at large couplings or small $\vrel$, independently of the partial wave~\cite{Cassel:2009wt,Petraki:2015hla}. Nevertheless, for a process that has a perturbative limit, the $\vrel^{2J}$ suppression of the higher partial waves that appears in the perturbative regime ($\vrel > \aD$), morphs into an $\aD^{2J}$ suppression in the Sommerfeld-enhanced regime ($\vrel < \aD$), as already evident in \cref{eqs:sigma_ann_ScalarMed}. This, in fact, happens also for higher-order corrections in $\vrel^2$ within a given partial wave~\cite{ElHedri:2016onc}. For such processes, it is typically true that the lowest non-vanishing partial wave yields the dominant contribution.

However, the coupling to a light mediator often implies a variety of radiative processes that can annihilate DM, some of which may not have a perturbative limit. In particular, bound-state formation is inherently non-perturbative. For such processes, the association between the partial-wave expansion and the $\vrel^2$ (or $\aD^2$, if $\aD>\vrel$) expansion does not hold.

For a vector mediator, the leading order contribution to the direct DM annihilation into radiation is dominantly $s$-wave,  while the radiative capture to the ground state is a $p$-wave process. Despite the different partial waves contributing, both the annihilation and BSF have the same dependence on $\aD$ for $\vrel\lesssim \aD$. Applying the limit \eqref{eq:sigmaUni} on \cref{eqs:VectorMed_sigmas}, using the appropriate value of $J$, we find that unitarity is violated by the leading order computation of $\s_\ann$ and $\sigma_\BSF$ at $\aD \gtrsim 0.85$; notably, this is approximately the same value of $\aD$ for both processes. Around and above this value of $\aD$, higher-order corrections must be included to accurately determine the cross-sections of interest. Nevertheless, it is evident that for such large values of $\aD$ 
--- around which the unitarity limit on the inelastic cross-sections may be realised --- 
the depletion of DM via $p$-wave inelastic scattering dominates over $s$-wave, for the velocity range that is relevant to the DM chemical decoupling in the early universe ($\vrel^\FO \lesssim 0.3$). It is also interesting that, around these values of $\aD$, the $s$-wave annihilation still gives a sizeable contribution (nearly the maximally allowed by unitarity), and $d$-wave inelastic scattering, which is the dominant mode of the capture into $n=2, \, \ell=1$ bound states, is comparable to the $s$-wave annihilation~\cite{Petraki:2016cnz} (without, though, saturating its unitarity limit).

Moreover, for a scalar mediator, the direct annihilation into radiation, which is the dominant inelastic interaction in the Coulomb regime, is a $p$-wave process at leading order, ${\cal M}_\ann \propto \cos \theta \propto d_{0,0}^1(\theta)$. In this case, the apparent violation of unitarity occurs for~$\aD \gtrsim 1.4$. It is possible that more complex models feature inelastic processes where even  higher partial waves dominate (see e.g.~Ref.~\cite{Asadi:2016ybp} for a model with a rich spectrum of radiative transitions).

From the above it is evident that higher partial waves are important. In \cref{sec:Unitarity_MassBounds}, we compute the $s$- and $p$-wave unitarity bounds, and their combination, on the mass of symmetric and asymmetric thermal-relic DM.

\subsection{Bounds on the mass of symmetric and asymmetric thermal-relic DM \label{sec:Unitarity_MassBounds}}

We first analytically estimate the unitarity bound on the DM mass. The DM relic density depends on the product $\lambda\Phi$, which in turn depends on the annihilation cross-section at freeze-out [cf.~\cref{eq:lambda,eq:Phi_approx}]. The thermal average of \cref{eq:sigmaUni} is $\< \sigma_\uni^{(J)} \vrel \>  = (2J+1) \, 4  \, (\pi \xD)^{1/2}/\MDM^2$. Then,
\beq
\lambda \Phi \ \leqslant \ \lambda_\uni \, \Phi_\uni 
\ = \
\frac{8\pi}{3} \sqrt{\frac{g_*^\FO}{5\xD^\FO}} \, \frac{\MPl}{\MDM} \times
\left\{
\bal{4}
&(2J+1),& \qquad  &\text{solely } J,& 
\\
&(J_{\max} +1)^2,& \qquad &0 \leqslant J \leqslant J_{\max},&
\eal
\right.
\label{eq:lambdaPhi_uni}
\eeq
depending on which partial waves contribute maximally to the DM annihilation.

For symmetric thermal-relic DM ($\epsilon = 0$), the relic density is [cf. \cref{eq:Yinf_sym}]
\beq
\OmegaDM \ = \ 
\frac{2 \MDM Y_\infty^\sym \, s_0 }{\rho_c} 
\ \simeq \ \frac{2 \MDM s_0 / \rho_c }{\lambda_\sym \Phi_\sym} 
\ \geqslant \ \frac{2\MDM s_0 / \rho_c }{\lambda_\uni \Phi_\uni} \,,
\label{eq:OmegaDM_sym_uni}
\eeq
where we omitted the factor $(1+c/\xD^\FO)^{-1} \simeq 1$.
This implies $\MDM \ \leqslant \ M_{\uni,\sym}$, with
\begin{subequations}
\beq
M_{\uni,\sym} =
M_{\uni,\sym}^{(0)} \times 
\left\{
\bal{4}
&\sqrt{2J+1},& \qquad  &\text{solely } J \,,& 
\\
&J_{\max} +1,& \qquad &0 \leqslant J \leqslant J_{\max} \,,&
\eal
\right.
\label{eq:Muni_sym}
\eeq
and\footnote{%
Note that $M_{\uni,\sym}^{(0)}$ has some sensitivity of on the assumptions about $\tilde{T}$ and the degrees of freedom in each sector, via $g_*^\FO$. Had we assumed that the dark plasma is at the same temperature as the SM plasma at the time of freeze-out, we would have found $M_{\uni,\sym}^{(J=0)} \approx 140$~TeV~\cite{vonHarling:2014kha}. According to the assumptions made here, for $\MDM \sim 100~\TeV$, the dark plasma is at a somewhat higher temperature than the SM at the time of DM freeze-out. This is due to the DM degrees of freedom becoming non-relativistic below the last common temperature $\tilde{T}$ of the two sectors, while no decoupling of SM degrees of freedom has yet occurred at $\TSM \sim 100~\TeV/\xD^\FO \sim \text{few }\TeV$. A hotter dark sector necessitates a larger DM annihilation cross-section, and therefore implies a stronger unitarity bound on $\MDM$. Note that these constraints can be considerably relaxed if there is an entropy injection into the thermal bath after DM freezeout~\cite{Bramante:2017obj}.}
%
\beq
M_{\uni,\sym}^{(0)} = 
\[ \frac{4\pi}{3} \ \sqrt{\frac{g_*^\FO }{5\xD^\FO} } 
\ \frac{\rho_c \MPl \OmegaDM}{s_0} \]^{1/2}
\approx 110~\TeV \,. 
\eeq 
\end{subequations}

Asymmetric thermal-relic DM requires a larger annihilation cross-section than symmetric DM. Unitarity, thus, sets a tighter upper bound on $\MDM$, that depends on the asymmetry $\epsilon = \etaD/\etaB$, 
$\MDM \leqslant M_\uni(\epsilon) \leqslant M_{\uni,\sym}$. 
Moreover, the DM mass is bounded from above by the value it would have if a vanishing late-time fractional asymmetry, $\rinf \to 0$, could be attained, 
$\MDM < M_{\max} (\epsilon) \simeq 5~\GeV/\epsilon$ [cf.~\cref{eq:MDM_max}]. 
$M_\uni(\epsilon)$ describes the transition between $M_{\uni,\sym}$ and $M_{\max} (\epsilon)$, which occurs at 
$M_{\rm max}(\epsilon) \sim M_{\uni,\sym}$, i.e.~for 
$\epsilon \sim 5~\GeV/M_{\uni,\sym} \sim 10^{-5}$. 
For $\epsilon \gg 10^{-5}$, the DM mass is bounded essentially by $M_{\rm max}(\epsilon)$ and very small $\rinf$ can be realised; 
for $\epsilon \ll 10^{-5}$, $\MDM$ is bounded by $M_{\uni,\sym}$ and $\rinf \sim {\cal O}(1)$.  
Conversely, the unitarity limit may be interpreted as a lower bound on the fractional asymmetry, $\rinf \geqslant r_{\infty,\uni}$; for small $\epsilon$ or large $\MDM$, $\rinf$ cannot be too small, which in turn may result in significant annihilation signals at late times.

\begin{figure}[t]
\centering
\includegraphics[width=0.48\textwidth]{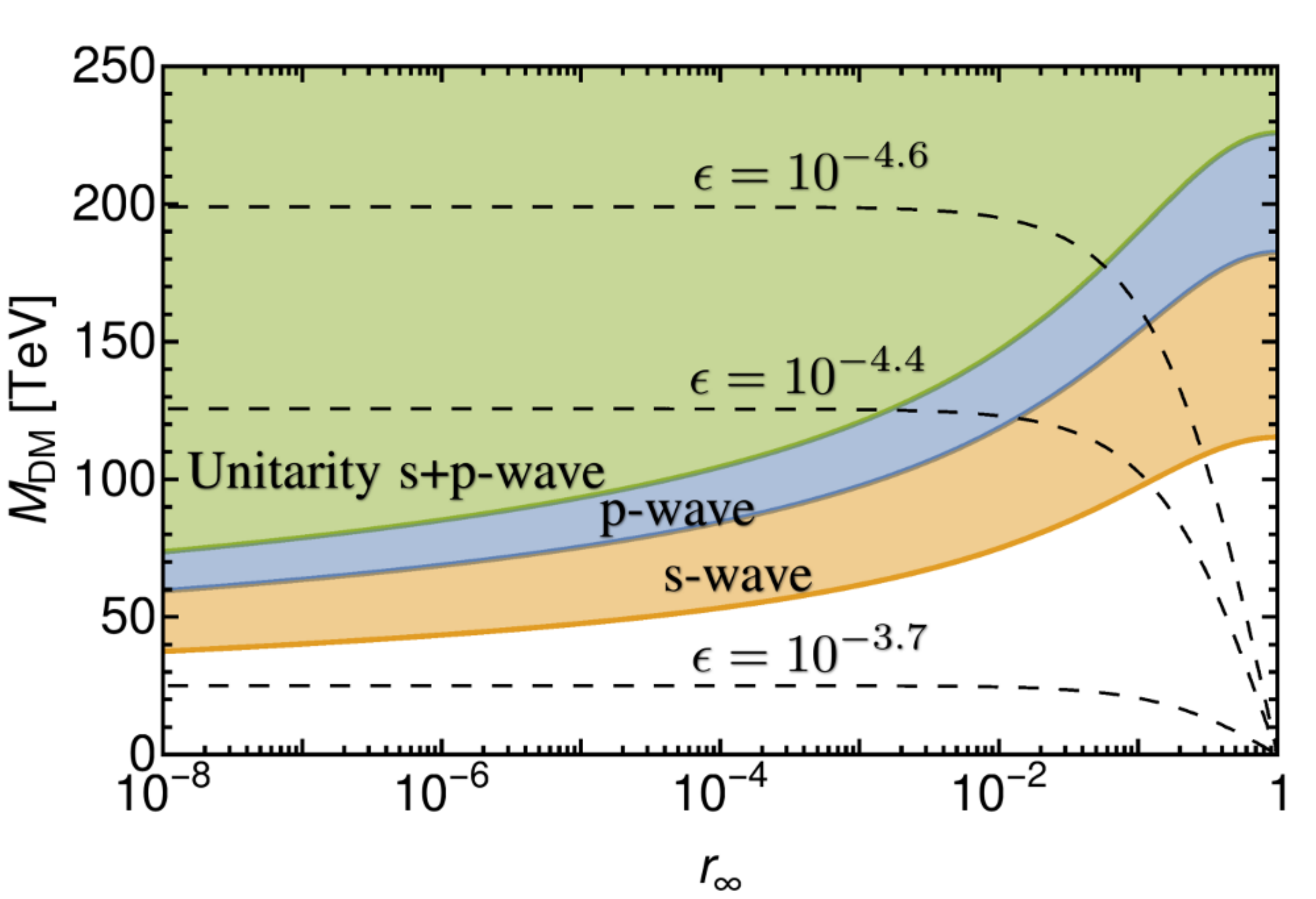}~~~
\includegraphics[width=0.48\textwidth]{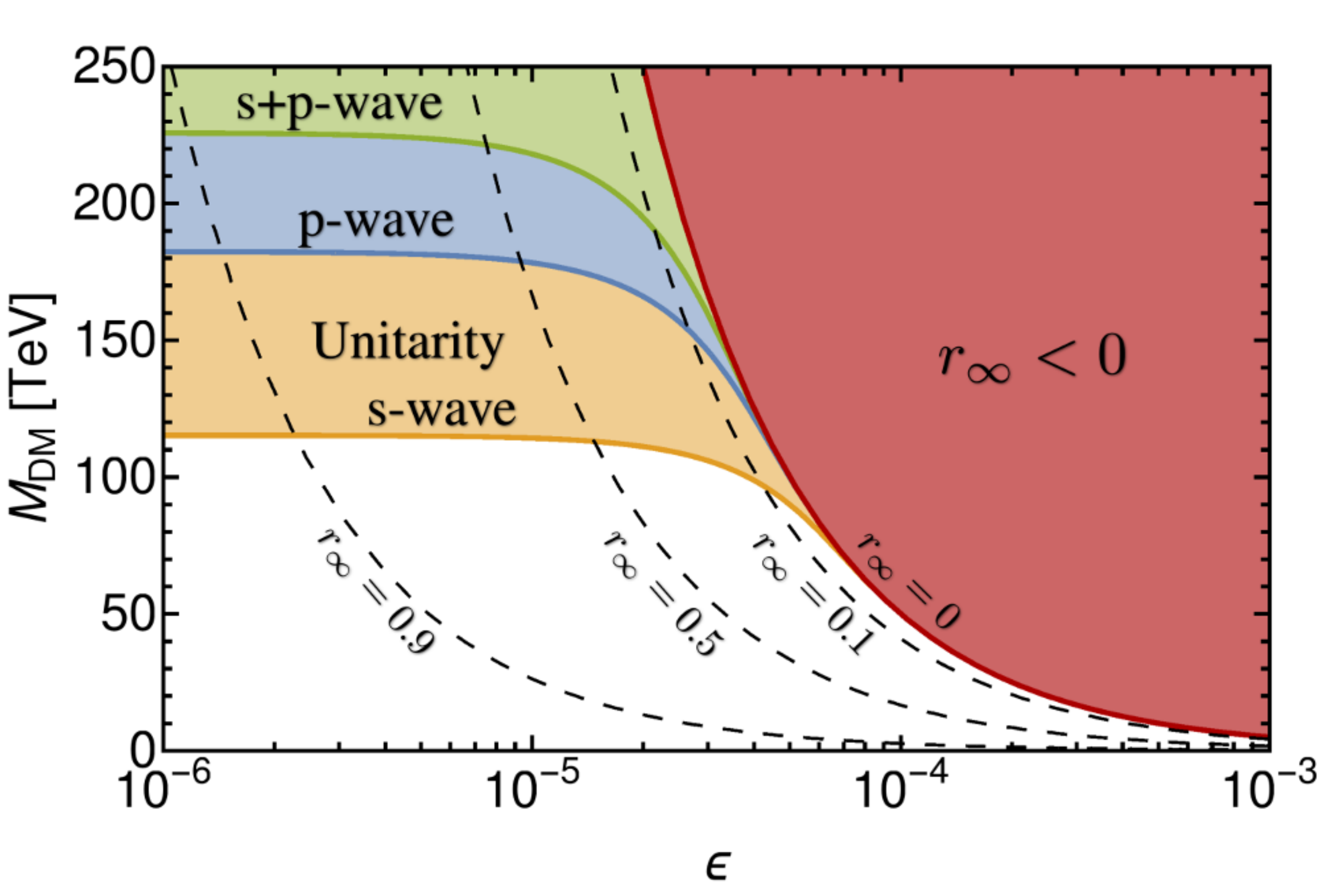}


\includegraphics[width=0.48\textwidth]{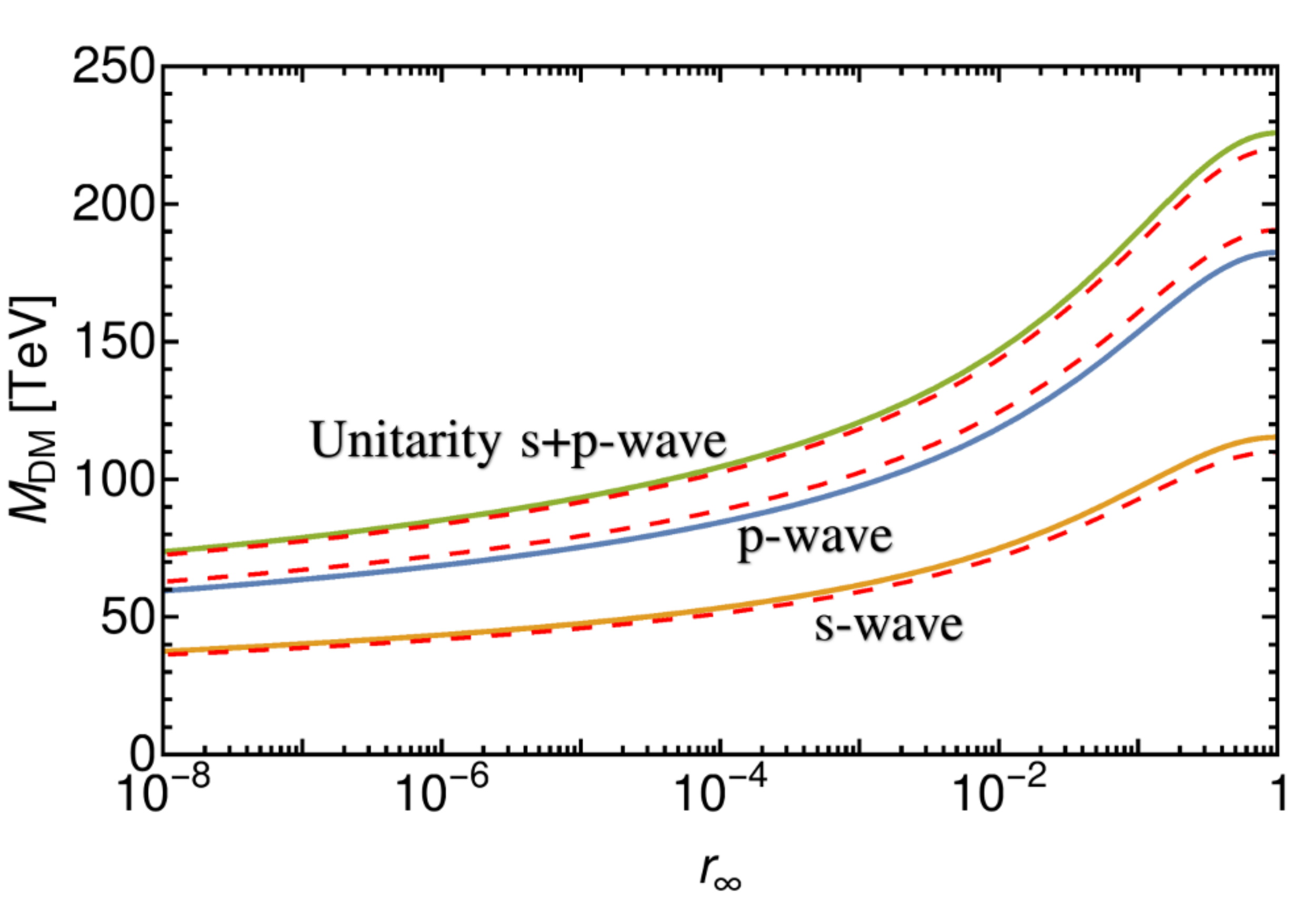}~~~
\includegraphics[width=0.48\textwidth]{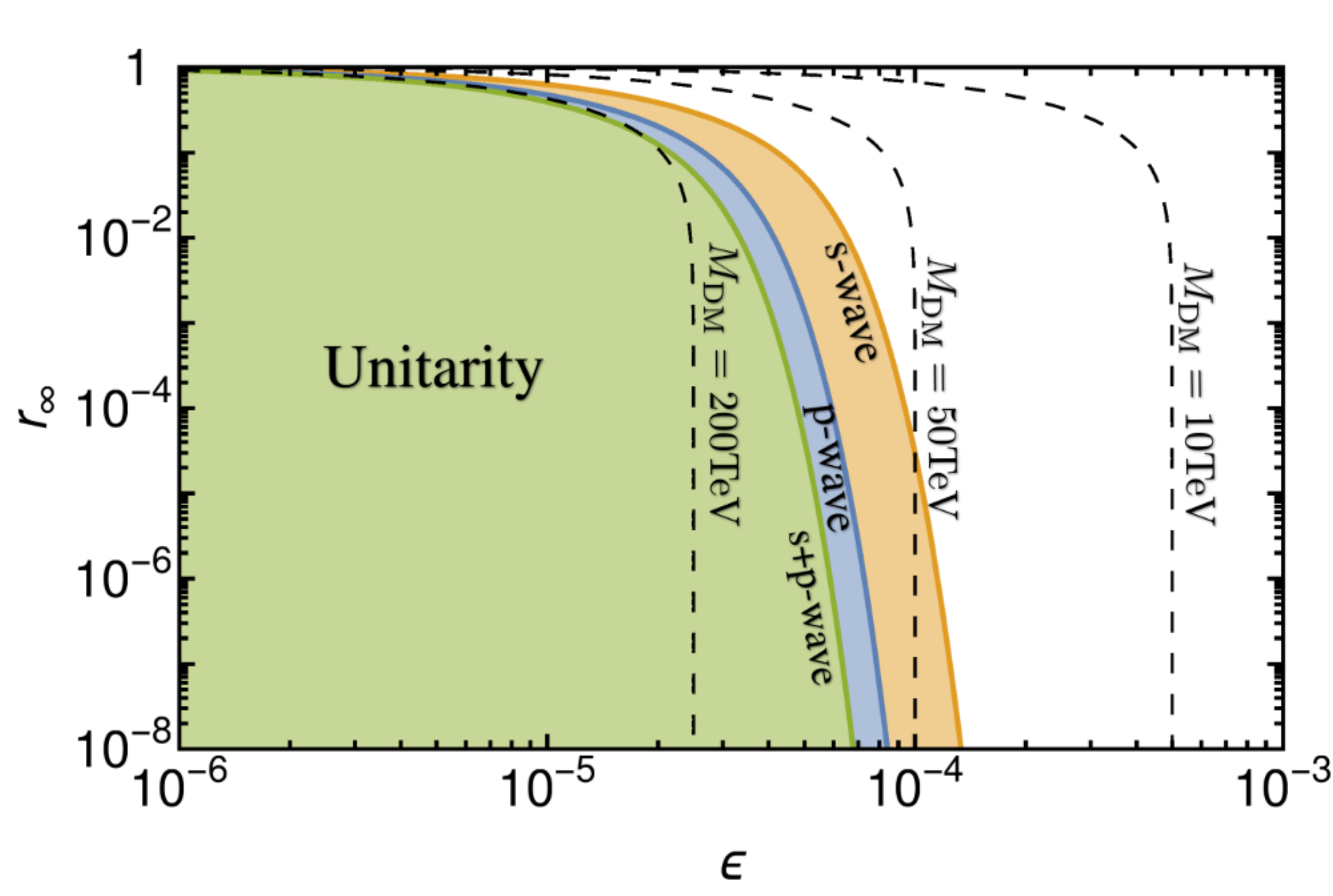}
\caption[]{\label{fig:Unitarity} 
Unitarity bounds on symmetric and asymmetric thermal-relic dark matter, annihilating via processes dominated by the $s$ or $p$ partial waves, or their sum. The $\epsilon$, $r_{\infty}$ and $M_{\rm DM}$ dashed contours illustrate the relation $M_{\rm DM} \simeq (5~{\rm GeV}/\epsilon) \times (1-r_{\infty})/(1+r_{\infty})$ [cf.~eq.~\eqref{eq:MDM}]. 
\emph{Top left:} For small values of the final fractional asymmetry $r_\infty$,  the bounds on $M_{\rm DM}^2$ tighten up approximately logarithmically with decreasing $r_\infty$.  
\emph{Top right:} At $\epsilon \equiv \eta_D/\eta_B \ll 10^{-5}$, the DM mass is bounded by the unitarity limit on symmetric thermal-relic DM, while at $\epsilon \gg 10^{-5}$, it is limited by $M_{\rm max} \simeq 5~{\rm GeV}/\epsilon$. 
\emph{Bottom right:} For $\epsilon \ll 10^{-5}$, DM retains a large symmetric component, $r_{\infty} \sim {\cal O}(1)$, while for $\epsilon \gg 10^{-5}$, small values of $r_{\infty}$ can be attained. 
\emph{Bottom left:} Comparison of numerical evaluation (solid) and analytical approximation (dashed) [cf.~eqs.~\eqref{eq:Muni_sym} and \eqref{eq:UniBounds_asym}].
}
\end{figure}

We may now estimate $M_\uni$ in the presence of an asymmetry. From \cref{eq:ln(rinf) expansion}, 
\beq
\ln \rinf 
\ \simeq  \ - \(1+ \frac{c}{\xD^\FO} \) \, \etaD \lambda \Phi
\ \gtrsim \ 
-\epsilon \etaB \lambda_\uni \Phi_\uni \,,
\label{eq:rinf_min}
\eeq
where $\lambda_\uni \Phi_\uni$ is given in \cref{eq:lambdaPhi_uni}, and $\epsilon$, $\rinf$ and $\MDM$ are also related via \cref{eq:MDM}. Solving \cref{eq:MDM} for $\epsilon$, $\rinf$ and $\MDM$, and substituting into \cref{eq:rinf_min}, we obtain respectively
\begin{subequations}
\label{eq:UniBounds_asym}
\label[pluralequation]{eqs:UniBounds_asym}
\beq
M_\uni \approx  M_{\uni,\sym} \times
\[ \(\frac{1+\rinf}{1-\rinf} \) \frac{\ln (1/\rinf)}{2} \]^{-1/2} \,,
\label{eq:Muni_vs_rinf}
\eeq
\beq
M_\uni \times 
\sqrt{\frac12 \(\frac{5~\GeV}{\epsilon \, M_\uni} \) 
\ln \[\frac{1+\epsilon \, M_\uni /(5~\GeV)}{1-\epsilon \, M_\uni /(5~\GeV)} \] }
\ \approx \ M_{\uni,\sym}  \,,
\label{eq:Muni_vs_epsilon}
\eeq
\beq
r_{\infty,\uni} =
\exp \[- 2 \, \(\frac{1+r_{\infty,\uni}}{1-r_{\infty,\uni}}\) \(\frac{\epsilon}{5~\GeV / M_{\uni,\sym}} \)^2 \] \,,
\label{eq:rinf_uni}
\eeq
\end{subequations}
where $M_{\uni,\sym}$ is given in \cref{eq:Muni_sym}. The last two equations can be solved numerically to obtain $M_\uni$ and $r_{\infty,\uni}$ in terms of $\epsilon$.

\Cref{eq:Muni_sym,eq:UniBounds_asym} provide an analytical approximation to the bounds implied by partial-wave unitarity, on symmetric and asymmetric thermal-relic DM. In \cref{fig:Unitarity}, we present the numerical computation of these bounds, for DM annihilation dominated by the $s$ and $p$ partial waves, or their sum.

\section{Conclusion \label{Sec:Concl}}

In a variety of models, asymmetric DM is hypothesised to couple to light force mediators. Models with hidden sectors are motivated either by high-energy physics, such as string-theory constructions (e.g.~\cite{Heckman:2011sw}), and/or on phenomenological grounds, for example by the similarity of the DM and ordinary matter densities (e.g.~\cite{Heckman:2011sw, Petraki:2011mv, vonHarling:2012yn, Choquette:2015mca}), 
self-interacting DM (e.g.~\cite{CyrRacine:2012fz, Cline:2013pca, Petraki:2014uza}) 
and dissipative DM~\cite{Foot:2014mia, Foot:2014uba, Foot:2015mqa, Fan:2013yva, Boddy:2016bbu, Agrawal:2017rvu}. 
Asymmetric DM coupled to light mediators may also consist of WIMPs 
with TeV-scale mass; indeed, for multi-TeV particles, the Weak interactions of the SM, mediated by $\sim 100~\GeV$ gauge bosons, manifest as long-range~\cite{Hisano:2002fk, Hisano:2003ec}.
In the present work, we focused on minimal models that feature long-range dynamics. We computed the DM freeze-out in the early universe, estimated the resulting annihilation signals at late times, and deduced constraints implied by unitarity.

Due to the Sommerfeld enhancement of the inelastic processes,  the couplings required to eliminate efficiently the dark antiparticles and establish a large final asymmetry ($\rinf \ll 1$), can be considerably lower than in the case of contact interactions. Within a specific model, this broadens the low-energy parameter space that yields highly asymmetric DM.

Despite lowering the predicted couplings, the Sommerfeld effect implies that  the indirect detection signals from the annihilations of the residual dark antiparticles can be significant. For example, for $\MDM \gtrsim \TeV$ and final fractional asymmetry as low as $\rinf \sim 10^{-3}$, the annihilation rate of asymmetric DM coupled to a light dark photon may be larger than that expected from symmetric DM with contact interactions. 
\Cref{fig:VectorMed_AnnSignals,fig:ScalarMed_AnnSignals} illustrate this point. 
This opens up the possibility of probing asymmetric DM with observations of the CMB, the Milky Way and the Dwarf galaxies. 
Moreover, the capture of DM in the interior of the Sun and its annihilation via metastable mediators can give rise to enhanced neutrino signals due to reduced absorption~\cite{Bell:2011sn}, and offers another opportune probe~\cite{Ardid:2015aol, Adrian-Martinez:2016ujo, Ardid:2017lry, Murase:2016nwx}. 
On the other hand, a sizeable DM annihilation rate may relax constraints from the 
capture of asymmetric DM in compact objects (see e.g.~\cite{Kouvaris:2011fi, Kouvaris:2011gb, Bramante:2013hn, Casanellas:2012jp, Bell:2013xk}).
However, we emphasise that the accurate determination of the expected annihilation signals necessitates that the entire cosmological history of any model of interest is first carefully considered; this may involve events such as the cosmological formation of stable bound states~\cite{Petraki:2014uza,Wise:2014jva}, that would suppress the annihilation signals estimated here. In fact, the radiative formation of stable bound states inside haloes is itself a potential source of indirect signals, albeit of lower energy~\cite{Pearce:2015zca,Pearce:2013ola}.

Long-range interactions, if adequately strong, can maximise the probability for inelastic scattering. In this regime of large inelasticity, higher partial waves can yield a significant, and in some cases the dominant contribution. However, even maximal inelasticity suffices to annihilate a thermal particle density down to the observed DM abundance only if these particles are not too heavy. The upper bound on the mass of thermal-relic DM implied by unitarity strengthens as the DM asymmetry increases. Conversely, unitarity implies that very heavy DM, or DM with small particle-minus-antiparticle-number-to-entropy ratio $\etaD$, has a significant symmetric component today.  The unitarity bounds are shown in \cref{fig:Unitarity}.

\section*{Acknowledgments}
We thank Adam Falkowski and Andreas Goudelis for very helpful discussions on unitarity, Sunny Vagnozzi for informative comments on $N_{\rm eff}$, Fazlollah Hajkarim for pointing out a subtlety regarding $g_*$,  and Kai Schmidt-Hoberg for comments on the manuscript.
K.P. was supported by the ANR ACHN 2015 grant (``TheIntricateDark" project), and by the NWO Vidi grant ``Self-interacting asymmetric dark matter".

\bibliography{Bibliography.bib}

\end{document}